\documentclass[pre,aps,preprint]{revtex4}
\pdfoutput = 1
\usepackage{graphicx}
\usepackage{amsfonts}
\usepackage{amssymb}
\usepackage{amsmath}

\usepackage{color}
\usepackage{psfrag}

\usepackage{epsfig}

\newcommand{\beq}{\begin{equation}}
\newcommand{\eeq}{\end{equation}}
\newcommand{\avg}[1]{\left< #1 \right>}
\newcommand{\bra}[1]{\big<#1|}
\newcommand{\ket}[1]{|#1\big>}
\newcommand{\braket}[2]{\big< #1 |#2\big>}
\newcommand{\barr}{\begin{eqnarray}}
\newcommand{\earr}{\end{eqnarray}}
\newcommand{\Ord}[1]{{\cal O}\left( #1\right)}

\newcommand{\ASd}[1]{}

\def\Paren#1{\left( #1 \right)}		

\def\id{\mathbb I}
\def\cD{{\cal D}}

\begin{document}

\title{Statistical properties of determinantal point processes in high-dimensional Euclidean spaces}

\author{Antonello Scardicchio}
\email{ascardic@princeton.edu}
\affiliation{Department of Physics, Joseph Henry Laboratories, Princeton University, Princeton, New Jersey 08544, USA; and
Princeton Center for Theoretical Science, Princeton University, Princeton, New Jersey 08544, USA}
\author{Chase E.~Zachary}
\email{czachary@princeton.edu}
\affiliation{Department of Chemistry, Princeton University, Princeton, New Jersey 08544, USA}
\author{Salvatore Torquato}
\email{torquato@princeton.edu}
\affiliation{Department of Chemistry, Princeton University, Princeton, New Jersey 08544, USA;\\
Program in Applied and Computational Mathematics, Princeton University, Princeton, New Jersey 08544, USA;\\
Princeton Institute for the Science and Technology of Materials, Princeton University, Princeton, New Jersey 08544, USA;\\
Princeton Center for Theoretical Science, Princeton University, Princeton, New Jersey 08544, USA; and
School of Natural Sciences, Institute for Advanced Study, Princeton, New Jersey 08540, USA}

\begin{abstract}

The goal of this paper is to quantitatively describe some statistical properties of higher-dimensional determinantal point processes with
a primary focus on the nearest-neighbor distribution functions.  
Toward this end, we express these functions as determinants of $N\times N$ matrices and then extrapolate to $N\rightarrow\infty$. This formulation
allows for a quick and accurate numerical evaluation of these quantities for point processes in Euclidean spaces of dimension $d$.  We also implement an algorithm due to Hough \emph{et.~al.} \cite{hough2006dpa} for generating configurations of determinantal point processes in arbitrary Euclidean spaces, and we utilize this algorithm in conjunction with the aforementioned 
numerical results to characterize the statistical properties of what we call the Fermi-sphere point process for $d = 1$ to $4$.  This homogeneous, isotropic determinantal point process, discussed also in a companion paper \cite{ToScZa08}, is the 
high-dimensional generalization of the distribution of eigenvalues on the unit circle of a random matrix from the circular unitary ensemble (CUE).
In addition to the nearest-neighbor probability distribution, we are able to calculate Voronoi cells and nearest-neighbor extrema statistics for the Fermi-sphere point process and discuss these as the dimension $d$ is varied.  The results in this paper accompany and complement analytical properties of higher-dimensional determinantal point processes developed in \cite{ToScZa08}.    

\end{abstract}

\maketitle

\section{Introduction}

Stochastic point processes (PPs) arise in several different areas of physics and mathematics. For example, 
the classical statistical mechanics of an ensemble of interacting point particles is essentially the study of a random point process 
with the Gibbs measure $d\mu(X) = P_N(X)dX=\exp[-\beta V(X)]dX$ 
providing the joint probability measure for an $N$-tuple of vectors $X=(\mathbf{x}_1,...,\mathbf{x}_N)$ to be chosen. Moreover, some many-body problems in quantum mechanics, 
as we will see, can be regarded as stochastic point processes, 
where quantum fluctuations are the source of randomness.  With regard to mathematical applications, it has been well-documented \cite{mehta2004rm} that
the distribution of zeros of the Riemann zeta function on the critical line is well-represented by the distribution of eigenvalues of a random $N\times N$ Hermitian matrix
from the Gaussian unitary ensemble (GUE) or circular unitary ensemble (CUE) in the limit $N\rightarrow\infty$.     
Nevertheless, it remains an open problem to devise efficient Monte Carlo routines aimed at sampling these processes in a computationally efficient way. 

In studies of the statistical mechanics of point-like particles one is usually interested in a handful of quantities
such as $n$-particle correlation functions,
the distributions of the spacings of particles, or the distributions of the sizes of cavities. 
Although these statistics involve only a small number of particles, it is not simple to extract them
from knowledge of the joint probability density $P_N$. 
In general numerical techniques are required because analytical results are rare. It is then of paramount importance 
to study point processes for which analytic results exist for at least some fundamental quantities. 
The quintessential example of such a process is the so-called Poisson PP, which is generated by placing points throughout the domain with a uniform probability distribution.
Such a process is completely uncorrelated and homogeneous, 
meaning each of the $n$-particle distribution functions is equal to $\rho^n$, where $\rho = N/V$ is the number density for the
process.  Configurations of points generated from this process are equivalent to classical systems of noninteracting particles or fully-penetrable 
spheres \cite{torquato2002rhm}, and almost all statistical descriptors may be evaluated analytically.

One nontrivial example of a family of processes which has been extensively studied is the class of determinantal PPs, 
introduced in 1975 by Macchi \cite{macchi1975cas} with reference to fermionic statistics.  
Since their introduction, determinantal point processes have found applications in diverse contexts, including random matrix theory (RMT), number theory, and physics 
(for a recent review, see \cite{soshnikov2000drp}). 
However, most progress has been possible in the case of point processes on the line and in the plane, 
where direct connections can be made with RMT \cite{mehta2004rm} and completely integrable systems \cite{jimbo1980dmi}. 

Similar connections have not yet been found, to the best of our knowledge, for higher dimensional determinantal point processes, 
and numerical and analytical results in dimension $d \geq 3$ are missing altogether. In this paper and its companion \cite{ToScZa08}, we provide a generalization of these point processes to higher dimensions which we call Fermi-sphere point processes. While in \cite{ToScZa08} we have studied, mainly by way of exact analyses, statistical descriptors such as $n$-particle
probability densities and nearest-neighbor functions for these point processes here we base most of our analysis on an efficient algorithm \cite{hough2006dpa} for generating configurations from arbitrary determinantal point processes, and are therefore able to study other particle and void statistics related to nearest-neighbor distributions and Voronoi cells.  
 
In particular, after presenting in detail our implementation of an algorithm \cite{hough2006dpa} to generate configurations of homogenous, isotropic determinantal point processes, we study several statistical quantities thereof, including Voronoi cells statistics and distributions of minimum and maximum nearest neighbor distances (for which no analytical results exist). Additionally, the large-$r$ behavior of the nearest-neighbor functions is computationally explored. We provide substantial evidence that the conditional probabilities $G_P$ and $G_V$, defined below, are asymptotically linear, and we give estimates for their slopes as a function of dimension $d$ between one and four. 

The plan of the paper is as follows. Section II provides a brief review of determinantal point processes and defines the statistical quantities used to characterize these systems.  Of particular 
importance is the formulation of the probability distribution functions governing nearest-neighbor statistics as determinants of $N\times N$ matrices; the results are 
easily evaluated numerically.  The terminology we develop is then applied to the statistical properties of known one- and two-dimensional determinantal point processes
in Section III.  Section IV discusses the implementation of an algorithm for generating determinantal point processes in any dimension $d$, and we combine the 
results from this algorithm and the numerics of Section II to characterize the so-called Fermi sphere point process for $d = 1, 2, 3,$ and $4$.  In Section V we provide an example of determinantal point-process on a curved space (a 2-sphere) and our conclusions are collected in Section VI.  

\section{Formalism of determinantal point processes}
\subsection{Definitions:  $n$-particle correlation functions}
Consider $N$ point particles in a subset of $d$-dimensional Euclidean space $E\subset \mathbb{R}^d$. It is convenient to introduce the Hilbert space structure given by square integrable functions on $E$; we will adopt Dirac's bra-ket notation for these functions. Unless otherwise specified, all integrals are intended to extend over $E$. A determinantal point process can be defined as a stochastic point process such that the joint probability distribution $P_N$
of $N$ points is given as a determinant of a positive, bounded operator 
$\mathcal{H}$ of rank $N$: 
\begin{equation}
\label{eq:probdef}
P_N(\mathbf{x}_1,...,\mathbf{x}_N)=\frac{1}{N!}\det[H(\mathbf{x}_i,\mathbf{x}_j)]_{1\leq i,j\leq N},
\end{equation}
where $H(\mathbf{x},\mathbf{y})$ is the kernel of $\mathcal{H}$. In this paper, we focus on the simple case in which the $N$ nonzero eigenvalues of $\mathcal{H}$ are all 1; 
the more general case can be treated with minor changes \cite{hough2006dpa}. We
can write down the spectral decomposition of $\mathcal{H}$ as:
\begin{equation}
\label{eq:defK}
\mathcal{H}=\sum_{n=1}^N\ket{\phi^0_n}\bra{\phi^0_n},
\end{equation}
where $\left\{\ket{\phi_n^0}\right\}_{n=1}^N$ are the eigenvectors of the operator $\mathcal{H}$.
The reason for the superscript on the basis vectors will be clarified momentarily. The correct normalization of the point process is obtained easily since \cite{mehta2004rm}:
\begin{equation}
  \int \det[H(\mathbf{x}_i,\mathbf{x}_j)]_{1\leq i,j\leq N}d\mathbf{x}_1\dotsm d\mathbf{x}_N=N!\det(\mathcal{H}),
\end{equation}
where the last determinant is to be interpreted as the product of the non-zero eigenvalues of the operator $\mathcal{H}$. Since these eigenvalues 
are all unity we obtain $\det(\mathcal{H})=1$, which yields:
\begin{equation}
\int P_N(\mathbf{x}_1,...,\mathbf{x}_N)d\mathbf{x}_1\dotsm d\mathbf{x}_N=1.
\end{equation}

Notice that in terms of the basis $\left\{\ket{\phi^0_n}\right\}_{n=1}^N$ we can also write:
\begin{equation}
\label{eq:probdefphi}
P_N(\mathbf{x}_1,...,\mathbf{x}_N)=\frac{1}{N!}\left|\det[\phi^0_i(\mathbf{x}_j)]_{1\leq i,j\leq N}\right|^2.
\end{equation}
An easy proof is obtained by considering the square matrix $\Phi_{ij}=\phi^0_i(\mathbf{x}_j)=\braket{\mathbf{x}_j}{\phi^0_i}$. Then,
\begin{eqnarray}
\left|\det[\phi^0_i(\mathbf{x}_j)]_{1\leq i,j\leq N}\right|^2&=&\det(\Phi^\dag)\det(\Phi)=\det(\Phi^\dag\Phi)\nonumber\\
&=&\det\left[\bra{\mathbf{x}_i}\left(\sum_{n=1}^N\ket{\phi^0_n}\bra{\phi^0_n}\right)\ket{\mathbf{x}_j}\right]=\det[H(\mathbf{x}_i,\mathbf{x}_j)],
\end{eqnarray}
which is the same as \eqref{eq:probdef}.

Determinantal point processes are peculiar in that one can actually write all the $n$-particle distribution functions explicitly. The \emph{$n$-particle probability density},
denoted by $\rho_n(\mathbf{x}_1,\dotsc,\mathbf{x}_n)$, is proportional to the probability density of finding the first $n$ particles in volume
elements around the given positions $(\mathbf{x}_1, \dotsc, \mathbf{x}_n)$, 
irrespective of the remaining $N-n$ particles. 
For a general determinantal point process this function takes the form:
\begin{equation}
\rho_{n}(\mathbf{x}_1,...,\mathbf{x}_{n})= \det[H(\mathbf{x}_i,\mathbf{x}_j)]_{1\leq i,j\leq n} = n!~P_n(\mathbf{x}_1,\dotsc,\mathbf{x}_n).
\end{equation}
In particular, the single-particle probability density is:
\begin{equation}
\rho_1(\mathbf{x}_1)=H(\mathbf{x}_1,\mathbf{x}_1).
\end{equation}
This function is proportional to 
the probability density of finding a particle at $\mathbf{x}_1$, also known as the \emph{intensity} of the point process. One can see that the normalization is: 
\begin{equation}
\int \rho_1(\mathbf{x}) d\mathbf{x}= \text{tr}(\mathcal{H})=N.
\end{equation}
For translationally-invariant processes $\rho_1(\mathbf{x})=\rho$, independent of $\mathbf{x}$. 
We remark in passing that for a finite system translational invariance is defined in the sense of averaging the location of the origin over $\mathbb{R}^d$ with 
periodic boundary conditions enforced.

It is also possible to write the two-particle probability density explicitly:
\begin{equation}\label{rho2}
\rho_2(\mathbf{x}_1,\mathbf{x}_2)= H(\mathbf{x}_1,\mathbf{x}_1)H(\mathbf{x}_2,\mathbf{x}_2)-\left\lvert H(\mathbf{x}_1,\mathbf{x}_2)\right\rvert^2,
\end{equation}
which has the following normalization:
\begin{equation}
\int \rho_2(\mathbf{x}_1,\mathbf{x}_2)d\mathbf{x}_1 d\mathbf{x}_2=N(N-1).
\end{equation}
In general the normalization for $\rho_n$ is given by $N!/(N-n)!$, or the number of ways of choosing an ordered subset of $n$ points from a population of size $N$.
For a translationally-invariant and completely uncorrelated point process \eqref{rho2} simplifies according to $\rho_2 = \rho^2$.

We also introduce the \emph{$n$-particle correlation functions} $g_n$, which are defined by:
\begin{equation}
g_n(\mathbf{x}_1,\dotsc,\mathbf{x}_n) = \frac{\rho_n(\mathbf{x}_1,\dotsc,\mathbf{x}_n)}{\rho^n}.
\end{equation}
Since $\rho_n = \rho^n$ for a completely uncorrelated point process, it follows that deviations of $g_n$ from unity provide a measure of the correlations
between points in a point process.
Of particular interest is the pair correlation function, which for a translationally-invariant point process of intensity $\rho$ can be written as:
\begin{equation}\label{g2detpp}
g_2(\mathbf{x}_1,\mathbf{x}_2)=\frac{\rho_2(\mathbf{x}_1,\mathbf{x}_2)}{\rho^2}=1-\left\lvert\frac{H(\mathbf{x}_1,\mathbf{x}_2)}{\rho}\right\rvert^2.
\end{equation}

Closely related to the pair correlation function is the \emph{total correlation function}, denoted by $h$; it is derived from $g_2$
via the equation:
\begin{equation}\label{hdef}
h(\mathbf{x},\mathbf{y}) = g_2(\mathbf{x},\mathbf{y}) - 1 = -\rho^{-2}\left\lvert H(\mathbf{x}, \mathbf{y})\right\rvert^2,
\end{equation}
where the second equality applies for all determinantal point processes by \eqref{g2detpp}.
Since $g_2(r) \rightarrow 1$ as $r\rightarrow \infty$ ($r = \lVert\mathbf{x}-\mathbf{y}\rVert$) for translationally invariant systems without long-range order, 
it follows that $h(r)\rightarrow 0$ in this limit, meaning that $h$ is generally 
an $L^2$ function, and its Fourier transform is well-defined.  

\ASd{It is common in statistical mechanics when passing to reciprocal space to
consider the associated \emph{structure factor} $S$, which for a translationally invariant system is defined by:
\begin{equation}\label{Sdef}
S(k) = 1+\rho\hat{h}(k),
\end{equation}
where $\hat{h}$ is the Fourier transform of the total correlation function, $\rho$ is the number density, and $k = \lVert\mathbf{k}\rVert$ is the magnitude of the 
reciprocal 
variable to $\mathbf{x}$.  
We utilize the following definition of the
Fourier transform:
\begin{equation}
\hat{f}(\mathbf{k}) = \int_{\mathbb{R}^d} f(\mathbf{x}) \exp\left[-i(\mathbf{k}, \mathbf{x})\right] d\mathbf{x},
\end{equation}
where $(\mathbf{k}, \mathbf{x}) = \sum_{i=1}^d k_i x_i$ is the conventional Euclidean inner product of two real-valued vectors.  
When it is well-defined, the corresponding inverse Fourier transform is given by:
\begin{equation}
f(\mathbf{r}) = \left(\frac{1}{2\pi}\right)^d \int_{\mathbb{R}^d} \hat{f}(\mathbf{k}) \exp\left[i(\mathbf{k}, \mathbf{x})\right] d\mathbf{k}.
\end{equation}
We remark that for radially-symmetric functions [i.e., $f(\mathbf{x}) = f(\lVert\mathbf{x}\rVert) = f(r)$], the Fourier and inverse Fourier transforms may
be written respectively as follows:
\begin{eqnarray}
\hat{f}(k) &=& (2\pi)^{d/2}\int_0^{\infty} r^{d-1} f(r) \frac{J_{(d/2)-1}(kr)}{(kr)^{(d/2)-1}} dr\\
f(r) &=& \left(\frac{1}{2\pi}\right)^{d/2} \int_0^{\infty} k^{d-1} \hat{f}(k) \frac{J_{(d/2)-1}(kr)}{(kr)^{(d/2)-1}} dk.
\end{eqnarray}
} 

Determinantal point processes are self-similar; integration of the $n$-particle probability distribution with respect to a point gives back the same functional form \footnote{One could think in terms of effective interactions and renormalization group. The determinantal form of the probabilities $\rho_n$ then is a fixed point of the renormalization operation of integrating out one or more particles.}. This property is desirable since it considerably simplifies the computation of many quantities. However, we note that even complete knowledge of all the $n$-particle probability distributions is not sufficient \emph{in practice} to generate point processes from the given probability $P_N$. This notoriously difficult issue is known as the reconstruction problem in statistical mechanics \cite{ToSt03, ToSt02, KuLeSp07, UcStTo06}. When in Section IV we discuss an explicit constructive algorithm to generate realizations of a given determinantal process, the reader should keep in mind that the ability to write down all the $n$-particle correlation functions $g_n$ is {\it not} the reason why there exists such a constructive algorithm.

\subsection{Exact results for some statistical quantities}

We have seen that the determinantal form of the probability density function allows us to write down all $n$-particle correlation functions $g_n$ in a quick and simple manner. However, we can also express more interesting functions, such as the probability of having an empty region $\cD$ or the expected number of points in a given region, as properly constructed determinants of the operator $\mathcal{H}$. This property has been used in random matrix theory to find the exact gap distribution of eigenvalues on the line in terms of solutions of a nonlinear differential equation \cite{tracy1994lsd}. The relevant formula is a special case of the result \cite{soshnikov2000drp} that the generating function of the distribution of the number points $n_{\cD}$ in the region $\cD$ is:
\begin{equation}
\avg{z^{n_{\cD}}}=\sum_{n\geq 0}P(n_{\cD}=n)z^n=\det[\id+(z-1)\chi_{\cD}\mathcal{H}\chi_{\cD}],
\end{equation}
where $\chi_{\cD}$ is the characteristic function of $\cD$, $\id$ is the identity operator, and $z\in\mathbb{R}$. 
We will also denote $P_n\equiv P(n_{\cD}=n)$. Therefore, the probability that
the region $\cD$ is empty is obtained by taking the limit $z\to 0$ in the previous formula.
The result is:
\begin{equation}
\label{eq:PD0}
P_0=\det\Paren{\id-\chi_{\cD}\mathcal{H}\chi_{\cD}}.
\end{equation}
Equation \eqref{eq:PD0} may be written more explicitly. Consider the eigenvalues $\lambda_i$ of $\chi_{\cD}\mathcal{H}\chi_{\cD}$. 
By the definition of the determinant, equation \eqref{eq:PD0} takes the form:
\begin{equation}
P_0=\prod_{i=1}^N(1-\lambda_i),
\end{equation}
where the product is over the non-zero eigenvalues of $\chi_{\cD}\mathcal{H}\chi_{\cD}$ only (of which there are $N$, the number of particles).
First notice that for the non-zero $\lambda_i$ we have $\lambda_i=\tilde{\lambda}_i$, where $\left\{\tilde{\lambda}_i\right\}_{i=1}^N$ are the 
$N$ eigenvalues of $\mathcal{H}\chi_{\cD}\mathcal{H}$. 
In fact one can show that the traces of all the powers of these two operators are the same using $\chi_{\cD}^2=\chi_{\cD},\ \mathcal{H}^2=\mathcal{H}$, 
and the cyclic property of 
the trace operation. This condition is sufficient for $N$ finite, and the limit $N\to \infty$ can be taken afterward. 
The operator $\mathcal{H}\chi_{\cD} \mathcal{H}$ can now be written in a basis $\{\phi_n\}_{n=1}^N$ as the matrix:
\begin{equation}\label{M}
M_{ij}({\cD})=\int_{{\cD}}\overline{\phi_i(\mathbf{x})}\phi_j(\mathbf{x}) d\mathbf{x},
\end{equation}
and the determinant in (\ref{eq:PD0}) as:
\begin{equation}
\label{eq:PD0M}
P_0=\det\Paren{\delta_{ij}-M_{ij}}.
\end{equation}
We will be using this formula often in the following analysis. 
\ASd{It is also not difficult to find the probability $P_1$ that $\cD$ contains exactly 1 particle 
by taking a derivative with respect to $z$ and then letting $z\to 0$; the result is:
\begin{equation}
P_1=P_0\avg{n_{\cD}}.
\end{equation}
}
The probability $P_0$ has a unique role in the study of various point processes \cite{torquato2002rhm}, in particular when $\cD = B(0; r)$, a ball of radius $r$ 
(for translationally-invariant processes the position of the center of the ball is immaterial). 
In this context, $P_0$ is called the \emph{void exclusion probability} $E_V(r)$ \cite{ToLuRu90, ToLe90, LuTo92, torquato2002rhm}, and we will adopt this name and notation in this paper (in \cite{ToScZa08} we have studied this quantity in an appropriate scaling limit, when $d\to\infty$).

However, there are statistical quantities of great importance which cannot be found with the above formalism. For example, one can examine the distribution of the
maximum or  minimum nearest neighbor distances in a determinantal point process, or the ``extremum statistics," and these quantities cannot be found easily by the above means. One could also explore the distribution of the Voronoi cell statistics or the percolation threshold for the PP. To determine these quantities we will have to rely on an explicit realization of a determinantal point process. The existence and the analysis of an algorithm to perform this task is a central topic of this paper.

We introduce now some quantities which characterize a PP \cite{ToLuRu90, ToLe90, LuTo92, torquato2002rhm}. We start with the above expression $E_V(r)$ for the probability of finding a spherical cavity of radius $r$ in the point process.  Analogously, one can define the probability of finding a spherical cavity of radius $r$ \emph{centered on a point of the process}, which we denote as $E_P(r)$. $E_P$ can be found in connection with $E_V$ using the following construction. Consider the probability of finding no points in the spherical shell of inner radius $\epsilon$ and outer radius $r$, which we call $E_V(r; \epsilon)$. This function can be obtained by either of the previous formulas (\ref{eq:PD0}) or (\ref{eq:PD0M}).  It is clear that $E_V(r)=E_V(r;0)$. It is also true that for sufficiently small $\epsilon$ the probability of having two or more points in the sphere of radius $\epsilon$ is negligibly small compared to the probability of having one particle. Hence, the probability $\Omega(r; \epsilon)$ of finding no particles in the spherical shell $B(0;r)\setminus B(0;\epsilon)$ conditioned on the presence of one point in a sphere of radius $\epsilon$ and volume $v(\epsilon)$ is: 
\begin{equation}
\Omega(r; \epsilon) = \frac{E_V(r;\epsilon)-E_V(r;0)}{\rho v(\epsilon)},
\end{equation}
and by taking the limit $\epsilon\to 0$ of this expression we find that:
\begin{equation}
E_P(r)=\lim_{\epsilon\to 0}\Omega(r; \epsilon).
\end{equation}
That $E_P(0)=1$ can be seen from the following argument: set $r=\epsilon+0^+$. Then, $E_V(\epsilon+0^+;\epsilon)=1$ because the region is infinitesimal and hence empty with probability 1, and $E_V(\epsilon;0)\simeq1- \rho v(\epsilon)$ since for sufficiently small $\epsilon$ we have at most one point in the region. One line of algebra provides the result.

Using this expression, we can derive an interesting and practical result for $E_P$. First, notice that $E_V(r;\epsilon)$ contains the matrix $M_{ij}(r;\epsilon)$ defined 
by \eqref{M},
which when $\epsilon\to 0$ becomes:
\begin{equation}
M_{ij}(r;\epsilon)\simeq M_{ij}(r)-v(\epsilon)\overline{\phi_i(0)}\phi_j(0).
\end{equation}
Moreover, if we assume that $\mathbb{I}-M$ is invertible, we can see that to first order in $\delta M$:
\begin{eqnarray}
\det(\mathbb{I}-M+\delta M)&=&\exp[\ln\det(\mathbb{I}-M+\delta M)]=\exp\{\text{tr}[\ln(\mathbb{I}-M+\delta M)]\}\nonumber\\
&\simeq&\exp\{\text{tr}[\ln(\mathbb{I}-M)]+\text{tr}[\delta M(\mathbb{I}-M)^{-1}]\} \nonumber\\
&\simeq&\det(\mathbb{I}-M)\{1+\text{tr}[\delta M(\mathbb{I}-M)^{-1}]\}\label{pert}.
\end{eqnarray}
From \eqref{pert} we find the final result:
\begin{equation}\label{EpEvdet}
E_P(r)=E_V(r)\ \text{tr}\left[A(\mathbb{I}-M)^{-1}\right],
\end{equation}
where $A_{ij}=\overline{\phi_i(0)}\phi_j(0)/\rho$. Notice that for $r\to 0$, we have $M\to 0$, and $E_P(0)=\text{tr}(A)=\sum_i|\phi_i(0)|^2/\rho=H(0,0)/\rho=1$ as expected.
 
These two primary functions can be used to define four other quantities of interest. Two are density functions:
\begin{eqnarray}
H_V(r)&=&-\frac{\partial E_V(r)}{\partial r}\label{Hvdef}\\
H_P(r)&=&-\frac{\partial E_P(r)}{\partial r}\label{Hpdef},
\end{eqnarray}
which can be interpreted as the probability densities of finding the closest particle at distance $r$ from a random point of the space 
or another random point of the process, respectively. 
The other two functions are conditional probabilities:
\begin{eqnarray}
G_V(r)&=&\frac{H_V(r)}{\rho s(r)E_V(r)}\label{Gvdef}\\
G_P(r)&=&\frac{H_P(r)}{\rho s(r)E_P(r)}\label{Gpdef},
\end{eqnarray}
which give the density of points around a spherical cavity centered respectively on a random point of the space or on a random point of the process. 
We note that $s(r)$ is the surface area of the $d$-dimensional sphere of radius $r$. 
We will study the behavior of these functions for some determinantal PPs in Sections III and IV of this paper.

From the definitions in \eqref{Hvdef}-\eqref{Gpdef} in conjuction with \eqref{eq:PD0M} and \eqref{EpEvdet}, it is possible to express $H_V, H_P, G_V,$ and $G_P$
as numerically-solvable operations on $N\times N$ matrices.  The results are:
\begin{eqnarray}
H_V(r) &=& E_V(r)~\text{tr}\left[(\mathbb{I}-M)^{-1}\frac{\partial M}{\partial r}\right]\\
H_P(r) &=& H_V(r)~\text{tr}\left[A(\mathbb{I}-M)^{-1}\right]\nonumber\\
&~&-E_V(r)~\text{tr}\left[A(\mathbb{I}-M)^{-1}\frac{\partial M}{\partial r}(\mathbb{I}-M)^{-1}\right]\\
G_V(r) &=& \left[\frac{1}{\rho s(r)}\right]\text{tr}\left[(\mathbb{I}-M)^{-1}\frac{\partial M}{\partial r}\right]\\
G_P(r) &=& G_V(r)-\left[\frac{1}{\rho s(r)}\right]\frac{\partial}{\partial r}\left\{\ln \text{tr}\left[A(\mathbb{I}-M)^{-1}\right]\right\}.\label{GpGvrel}
\end{eqnarray}
The form $G_P(r) = G_V(r) - \tilde{G}(r)$ (which serves as a definition of $\tilde{G}$) in \eqref{GpGvrel} is of particular interest.  If the correction term $\tilde{G}(r) \geq 0$ for all $r$,
positivity and monotonicity of $G_P$ (which must be proven independently) are then sufficient to ensure that for appropriately large $r$, $G_P(r) \sim G_V(r)$ in scaling.  
Although we have been unable to develop analytic results for the large-$r$
behavior of $\tilde{G}$, numerical results, which are provided later (see Fig.\ \ref{Gtilde}), suggest that $\tilde{G} > 0$ and $\tilde{G}\rightarrow 0$ monotonically as $r \rightarrow \infty$ for $d \geq 2$, and $\tilde{G}\rightarrow$ constant for $d = 1$.  As both behaviors are subdominant with respect to the linear growth of $G_V$, we expect that $G_P$ and $G_V$ possess the same linear slope for sufficiently large $r$. 

\begin{figure}[!htbp]
\centering
\includegraphics[width=0.45\textwidth]{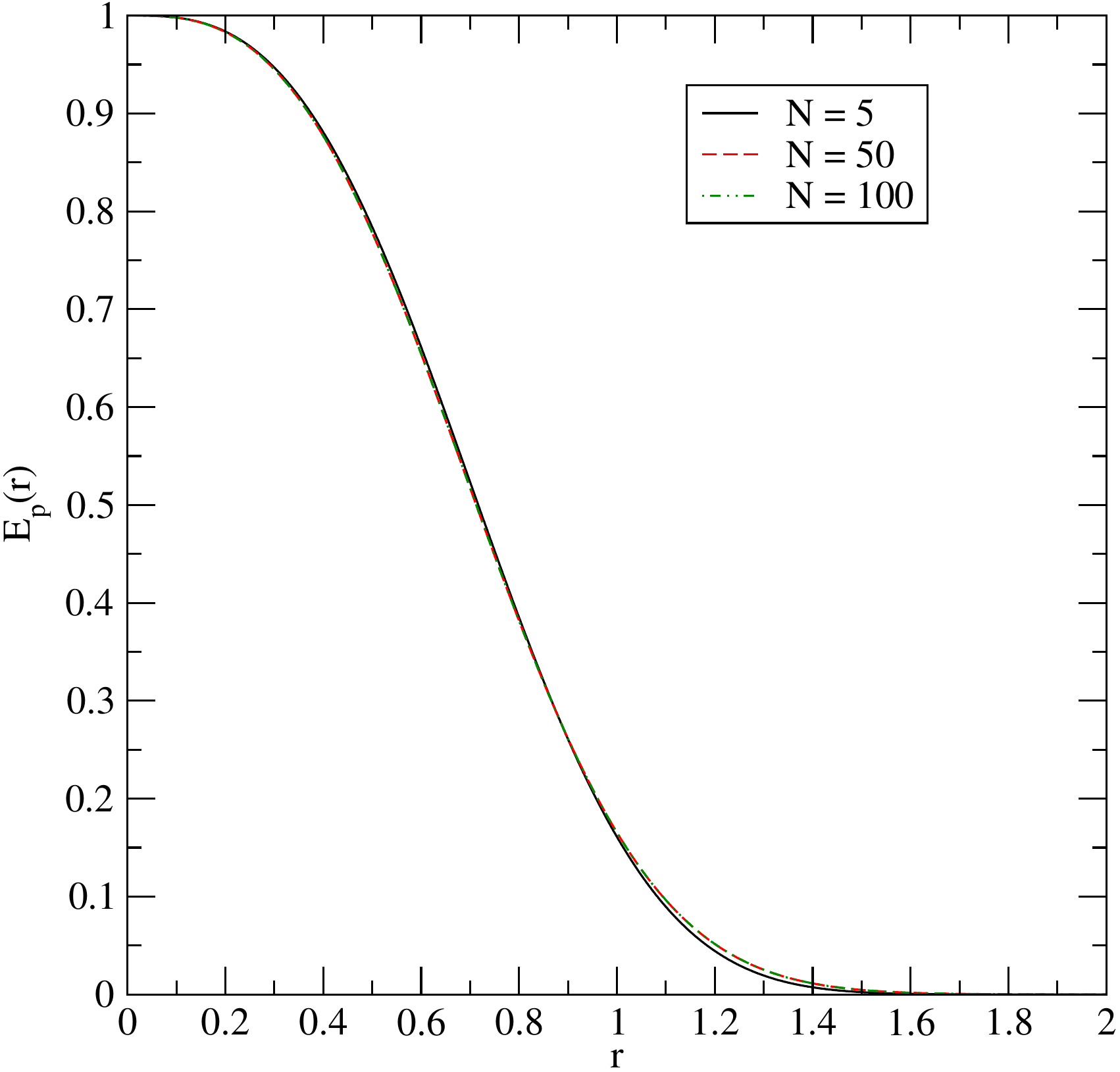}
\caption{Convergence of the $d = 1$ numerical results using \eqref{eq:PD0M} for $E_P$ with respect to increasing matrix size $N$.}\label{1dEpconv}
\end{figure}

An important point to address is the convergence of the results from \eqref{eq:PD0M} in the limit $N\rightarrow \infty$.  We expect that the calculations for finite
but large
$N$ provide an increasingly sharp approximation to the results from the $N\rightarrow \infty$ limit.  Figure \ref{1dEpconv} presents the
calculation of $E_P$ for a few values of $N$ with $d = 1$; 
it is clear that the numerical calculations quickly approach a fixed function for $N\gtrsim 40$, and it is this function
which we accept as the correct large-$N$ limit.  The results for higher-dimensional processes are similar, 
and we will assume that this convergence property holds throughout the 
remainder of the paper.

\subsection{Hyperuniformity of point processes}

Of particular significance in understanding the properties of determinantal point processes is the notion of
\emph{hyperuniformity}, also known as superhomogeneity.
A hyperuniform point pattern is a system of points such that the variance $\sigma^2(R) = \langle N_R^2\rangle - \langle N_R\rangle^2$ 
of the number of points $N_R$ in a spherical window of radius $R$ obeys:
\begin{equation}
\sigma^2(R) \sim R^{d-1}
\end{equation}
for large $R$ \cite{ToSt03}.  This condition in turn implies that the structure factor $S(\mathbf{k}) = 1+\rho\hat{h}(\mathbf{k})$ 
has the following small-$\mathbf{k}$ behavior:
\begin{equation}\label{Skhyper}
\lim_{\lVert\mathbf{k}\rVert\rightarrow\mathbf{0}} S(\mathbf{k}) = 0,
\end{equation}
meaning that hyperuniform point patterns do not possess infinite-wavelength number fluctuations \cite{ToSt03}.  Examples of hyperuniform systems
include all periodic point processes \cite{ToSt03}, certain aperiodic point processes \cite{ToSt03, GaJaJoLe03}, one-component plasmas \cite{ToSt03, GaJaJoLe03},
point processes associated with a wide class of tilings of space \cite{GaTo04, GaJoTo08}, and certain disordered sphere packings \cite{ToSt06b, ToScZa08, ToSt02, ToSt02a}.
It has also been shown \cite{ToScZa08} that the Fermi sphere determinantal point process, described below, is hyperuniform.

The condition in \eqref{Skhyper} suggests that for general translationally invariant nonperiodic systems:
\begin{equation}\label{Sscale}
S(k) \sim k^{\alpha}\qquad (k\rightarrow 0)
\end{equation}
for some $\alpha > 0$.  
However, hyperuniform determinantal point processes may only exhibit certain scaling exponents $\alpha$. \ASd{to prove this observation, we utilize the definition of
$S(k)$ in \eqref{Sdef} along with the relation between the total correlation function $h$ and the Hermitian kernel $H$ in \eqref{hdef}, which imply:} One can see for 
a determinantal point process that:
\begin{equation}\label{Sdetpphyp}
S(k) = 1-\rho\mathfrak{F}\{\lvert H\rvert^2\}(k),
\end{equation}
where $\mathfrak{F}$ denotes the Fourier transform.  Equation \eqref{Sdetpphyp} therefore suggests that:
\begin{equation}\label{FTinterm}
\mathfrak{F}\{\lvert H \rvert^2\}(k) \sim \left(\frac{1}{\rho}\right)\left(1-k^{\alpha}\right)\qquad (k\rightarrow 0).
\end{equation}    
Taking the inverse Fourier transform of \eqref{FTinterm} gives the following large-$r$ scaling of $\lvert H(r)\rvert^2$:
\begin{equation}\label{Hrscale}
\lvert H(r)\rvert^2 \sim -\left(\frac{1}{\rho\pi^{d/2}}\right)\left[\frac{2^{\alpha}~\Gamma(\frac{\alpha+d}{2})}{r^{\alpha+d}~\Gamma(-\frac{\alpha}{2})}\right]
\qquad (r\rightarrow \infty).
\end{equation}
The negative coefficient and the negative argument of the gamma function in \eqref{Hrscale} are crucially important.  Since $\left\lvert H(r)\right\rvert^2 \geq 0$
for all $r$, it must be true $\Gamma(-\alpha/2) < 0$, and this condition restricts the possible values of the scaling exponent $\alpha$.  Namely, the behavior
of the gamma function requires that $\alpha$ fall into one of the intervals $(0, 2]$, $[4, 6]$, $[8, 10]$, and so forth.  We remark that the integer-valued endpoints of these
intervals are indeed valid choices for $\alpha$ and imply that $\left\lvert H(r)\right\rvert \sim 0$ for sufficiently large $r$.  These values of $\alpha$ are therefore 
types of ``limiting values'' that
overcome the otherwise dominant $r^{-(\alpha+d)}$ asymptotic scaling of $\left\lvert H(r)\right\rvert^2$.  We provide an example of a determinantal point process with 
the critical scaling $\alpha = 2$ in Section III.C; the resulting large-$r$ behavior for $H(r)$ is seen to be exponential.  

\ASd{It is important to recognize that the aforementioned analysis applies specifically to determinantal point processes and is not a general result of hyperuniform point 
patterns.  The positivity of \eqref{Hrscale} is a direct result of the requirement that the system be negatively correlated, meaning
$h(r) \leq 0$ for all $r$; for an arbitrary hyperuniform point pattern, it is unclear that this behavior
must continue to hold.  Indeed the ``forbidden'' values of $\alpha$ for determinantal point processes are ``allowed'' values for a related class of 
point processes known as \emph{permanental point processes}, where the $n$-particle distribution functions are related to permanents of a 
positive semidefinite self-adjoint operator \cite{hough2006dpa}. This conclusion follows from an analysis similar to the one presented above.
More generally, it is not difficult to posit a form of $h(r)$ which exhibits ``forbidden'' asymptotic scaling; unfortunately,
there is no guarantee that such a function is realizable as the total correlation function of a classical system of $N$ interacting particles 
\cite{ToSt03, ToSt02, KuLeSp07, UcStTo06}.  
However, so long as $h(r) \geq -1$ and $S(k) \geq 0$ for the proposed function, we can at least claim that the hypothetical system is not
unreasonable.}       

\section{Properties of known determinantal point processes}

\subsection{One-dimensional processes}

By far, the most widely studied examples of determinantal point processes are in one-dimension. In fact, the connection to (RMT)
led others to explore the statistical properties of these systems even prior to the formal introduction of determinantal point processes.
To make this connection explicit, consider an $N \times N$ random Gaussian Hermitian matrix, i.e., a matrix whose elements
are independent random numbers distributed according to a normal distribution. 
This class of matrices defines the \emph{Gaussian unitary ensemble} or GUE. It is possible to see \cite{mehta2004rm} that the distribution induced on the eigenvalues of these random matrices is: 
\begin{equation}
\label{eq:GUEdist}
\rho_N(\lambda_1,...,\lambda_N)=\frac{1}{Z_N}\prod_{i<j}|\lambda_i-\lambda_j|^2 \exp\left[-\sum_i \lambda_i^2\right],
\end{equation}
where $Z_N$ is an appropriate normalization constant.
By a standard identity for the Vandermonde determinant:
\begin{equation}
\prod_{i<j}|\lambda_i-\lambda_j|^2=|\det(\lambda_i^n)_{1\leq i,n\leq N}|^2,
\end{equation}
and by combining the rows of the matrix $\lambda_i^n$ appropriately we find:
\begin{equation}
\prod_{i<j}|\lambda_i-\lambda_j|^2=|\det[H_j(\lambda_i)]_{1\leq i,j\leq N}|^2,
\end{equation}
where the functions $H_n(x)$ are the Hermite orthogonal polynomials normalized such that the coefficient of the highest power $x^n$ of $H_n$ is unity.
Taking into account the weight $e^{-x^2}$, we can write in agreement with (\ref{eq:probdefphi}):
\begin{equation}
p_N=\frac{1}{N!}|\det[\phi_j(\lambda_i)]_{1\leq i<j\leq N}|^2,
\end{equation}
where the orthonormal basis set is:
\begin{equation}
\phi_n(x)=\frac{1}{\sqrt{z_n}}H_n(x)\exp[-x^2/2];
\end{equation}
$z_n$ is a normalization factor. Therefore, this distribution is equivalent to the the one induced by a system of non-interacting, spinless fermions in a harmonic potential. 
We note without proof that the other canonical random matrix ensembles (GOE and GSE) can also be expressed as determinantal point processes by introducing an internal vector index for the basis functions \cite{mehta2004rm, soshnikov2000drp}. 

Another prominent example of a $d = 1$ determinantal point process is given by the unitary matrices distributed according to the invariant Haar measure; the resulting class is
termed the \emph{circular unitary ensemble} or CUE \cite{mezzadri2006grm}. The eigenvalues of these matrices can be written 
in the form $\lambda_j=e^{i\theta_j}$ with $\theta_j\in[0,2\pi]$ 
$\forall j \in \mathbb{N}$; they are distributed according to (\ref{eq:probdefphi}) with the basis:
\begin{equation}
\phi_n(\theta)=\frac{1}{\sqrt{2\pi}}\exp[in\theta].
\end{equation}
Notice that the eigenvalues represent the positions of free fermions on a circle, where the Fermi sphere has been filled continuously from momentum 0 to $N-1$. 

Another possible one-dimensional process is obtained by changing the exponent $x^2$ in (\ref{eq:GUEdist}) to an arbitrary polynomial. This generalization has interesting connections to the combinatorics of Feynman diagrams and to random polygonizations of surfaces \cite{difrancesco1995dga}. For other examples of one-dimensional determinantal point processes we refer the reader to \cite{soshnikov2000drp}.

\subsection{Exact results in one dimension}

For historical reasons, the most-studied descriptor of determinantal point processes is the gap distribution function,
which represents the probability density of finding a chord of length $s$ separating two points in the system for $d = 1$;
we denote this function by $p(s)$. For canonical ensembles of random matrices exact solutions for  $p(s)$ have been written in terms of solutions of well-known nonlinear differential equations \cite{mehta2004rm}.
We start with the following observation: after an appropriate rescaling of the eigenvalues, the gap distribution of eigenvalues of a random matrix is a universal function, depending only on the ``nature" of the ensemble (unitary, orthogonal or symplectic) which defines the small-$r$ behavior of $g_2$. For example, the two ensembles GUE and CUE defined above will have the same gap distribution in the limit $N\to \infty$. In the case of the GUE the limit is taken for the eigenvalues:
\begin{equation}
\lambda_i=z+\frac{\pi}{\sqrt{2N}}y_i,
\end{equation}
where $z$ is in the ``bulk" of the distribution ($|z|<\sqrt{2N}-\epsilon$ for $N$ large). One can prove that all the eigenvalues of a large random matrix will fall in an interval of size $2\sqrt{2N}$ with probability 1 in the large-$N$ limit. After this rescaling, the kernel $H$ converges to the ``sine kernel" in the large-$N$ limit \cite{tracy1994lsd,tracy1994fdd}:
\begin{equation}
H_N(\lambda_1,\lambda_2)\underset{N\rightarrow \infty}{\longrightarrow}H(y_1,y_2)=\frac{\sin[\pi(y_1-y_2)]}{\pi(y_1-y_2)}.
\end{equation}
From this result one can find the $n$-particle correlation functions. In particular one finds for $g_2$:
\begin{equation}
g_2(x,y)=1-\left(\frac{\sin[\pi(x-y)]}{\pi(x-y)}\right)^2.
\end{equation}

Applying this procedure to the CUE leads to the very same kernel; for a wider class of examples relevant to physics see \cite{brezin1993ucb}. 
Convergence of the kernel implies weak convergence of all the $n$-particle correlation functions to universal distributions. These distributions are defined by the sine kernel, one of a small family of kernels which appear to be universal \cite{tracy1994lsd,tracy1994fdd} in controlling large-$N$ limits of various statistical quantities of apparently different distributions. The study of the analytic properties of the kernels in this family yields a complete solution for the Janossy probabilities and edge distributions in one-dimensional systems.

Once the limiting kernel is identified, a solution for the gap distribution $p(s)$ still requires a detailed mathematical analysis \cite{tracy1994lsd}. 
An approximate form for $p(s)$, known as Wigner's surmise, was suggested by Wigner in 1951:
\begin{equation}
p(s)=\frac{32s^2}{\pi^2} \exp\left[{-\frac{4s^2}{\pi}}\right],
\end{equation}
and it is an extremely good fit for numerical data.
\ASd{; note that the density is normalized to one particle per unit length. The exact expression \cite{tracy1994lsd} can be rewritten in an analogous form \cite{forrester2000ews}:
\begin{equation}
p(s)=-\frac{\sigma(\pi s)}{s}\exp\left[\int_0^{\pi s} \frac{\sigma(t)}{t} dt\right],
\end{equation}
where $\sigma$ is the solution of a Painlev\'e V equation:
\begin{equation}
\label{eq:PV}
s^2(\sigma'')^2+4(s\sigma'-\sigma)[s\sigma'-\sigma+(\sigma')^2]-4(\sigma')^2=0.
\end{equation}
In the limit $s\rightarrow 0$, the solution scales according to:
\begin{equation}
\label{eq:PVbc}
\sigma(s)=-\frac{s^3}{3\pi}+\Ord{s^4}.
\end{equation}
An asymptotic analysis of \eqref{eq:PV} allows one to determine the large- and small-$s$ scalings of $p(s)$ and its related functions exactly.}  However, our
primary focus in this work is on the asymptotic behavior of the conditional probability $G_V$, and we therefore look for an exact solution for this function.      
\ASd{To characterize the large-$s$ behavior of this function,} First, 
we note without proof \cite{forrester2000ews} that $E_V(s)$ for $d = 1$ may be expressed in terms of a Painleve V transcendent.  
Namely, let $\tilde{\sigma}(s)$ be a solution of the nonlinear equation:
\begin{equation}\label{PV}
(s\tilde{\sigma}^{\prime\prime})^2+4(s\tilde{\sigma}^{\prime}-\tilde{\sigma})\left[s\tilde{\sigma}^{\prime}-\tilde{\sigma}+(\tilde{\sigma}^{\prime})^2\right]=0,
\end{equation}
subject to the boundary condition:
\begin{equation}
\tilde{\sigma}(s) \sim -\frac{s}{\pi}-\left(\frac{s}{\pi}\right)^2
\end{equation}
as $s\rightarrow 0$.  We may then write $E_V(s)$ in the form:
\begin{equation}\label{E2sigma}
E_V(s) = \exp\left\{\int_0^{2\pi s} \left[\frac{\tilde{\sigma}(t)}{t}\right]dt\right\}.
\end{equation}

We recall that $E_V(s)$ may also be expressed in terms of $G_V$ via the following relation:
\begin{equation}\label{EvGv}
E_V(s) = \exp\left[-2\int_0^s G_V(x) dx\right].
\end{equation}
By making a change of variables and comparing \eqref{E2sigma} and \eqref{EvGv}, we conclude that:
\begin{equation}\label{Gvsigma}
G_V(s) = -\frac{\tilde{\sigma}(2\pi s)}{2s}.
\end{equation}
Equation \eqref{Gvsigma} allows us to develop small- and large-$s$ expansions of $G_V$ in terms of
the equivalent expansions for $\tilde{\sigma}$. 

To describe the small-$s$ behavior of $G_V$, we substitute an expansion of the form:
\begin{equation}\label{sigmasmalls}
\tilde{\sigma}(s) = -\frac{s}{\pi} - \left(\frac{s}{\pi}\right)^2+\sum_{n=3}^{N} b_n s^n
\end{equation}
into \eqref{PV} and solve order-by-order for the coefficients $b_n$.  Upon converting the solution to a result for $G_V$ using \eqref{Gvsigma}, we obtain:
\begin{align}\label{Gvsmalls}
G_V(s) &= 1+2s+4s^2+\left(8-\frac{8\pi^2}{9}\right)s^3+\left(16-\frac{20\pi^2}{9}\right)s^4+\left(32-\frac{16\pi^2}{3}+\frac{64\pi^4}{225}\right)s^5\nonumber\\
&+\left(64-\frac{112\pi^2}{9}+\frac{448\pi^4}{675}\right)s^6+\mathcal{O}(s^7).
\end{align}

The derivation of the large-$s$ expansion is similar.  We choose an expansion of the form:
\begin{equation}\label{sigmalarges}
\tilde{\sigma}(s) = b_0s^2+b_1s+b_2+\sum_{n=3}^{N}b_ns^{2-n}
\end{equation}
and substitute this equation into \eqref{PV}.  After converting the result to an asymptotic series for $G_V$ with \eqref{Gvsigma}, we obtain:
\begin{equation}\label{Gvlarges}
G_V(s) = \frac{\pi^2s}{2}+\frac{1}{8s}+\frac{1}{32\pi^2s^3}+\frac{5}{64\pi^4s^5}+\frac{131}{256\pi^6s^7}+\frac{6575}{1024\pi^8s^9}+\frac{1080091}{8192\pi^{10}s^{11}}+\frac{16483607}{4096\pi^{12}s^{13}}+\mathcal{O}(s^{-15}).
\end{equation}

\begin{figure}[!tbp]
\centering
\includegraphics[width=0.45\textwidth]{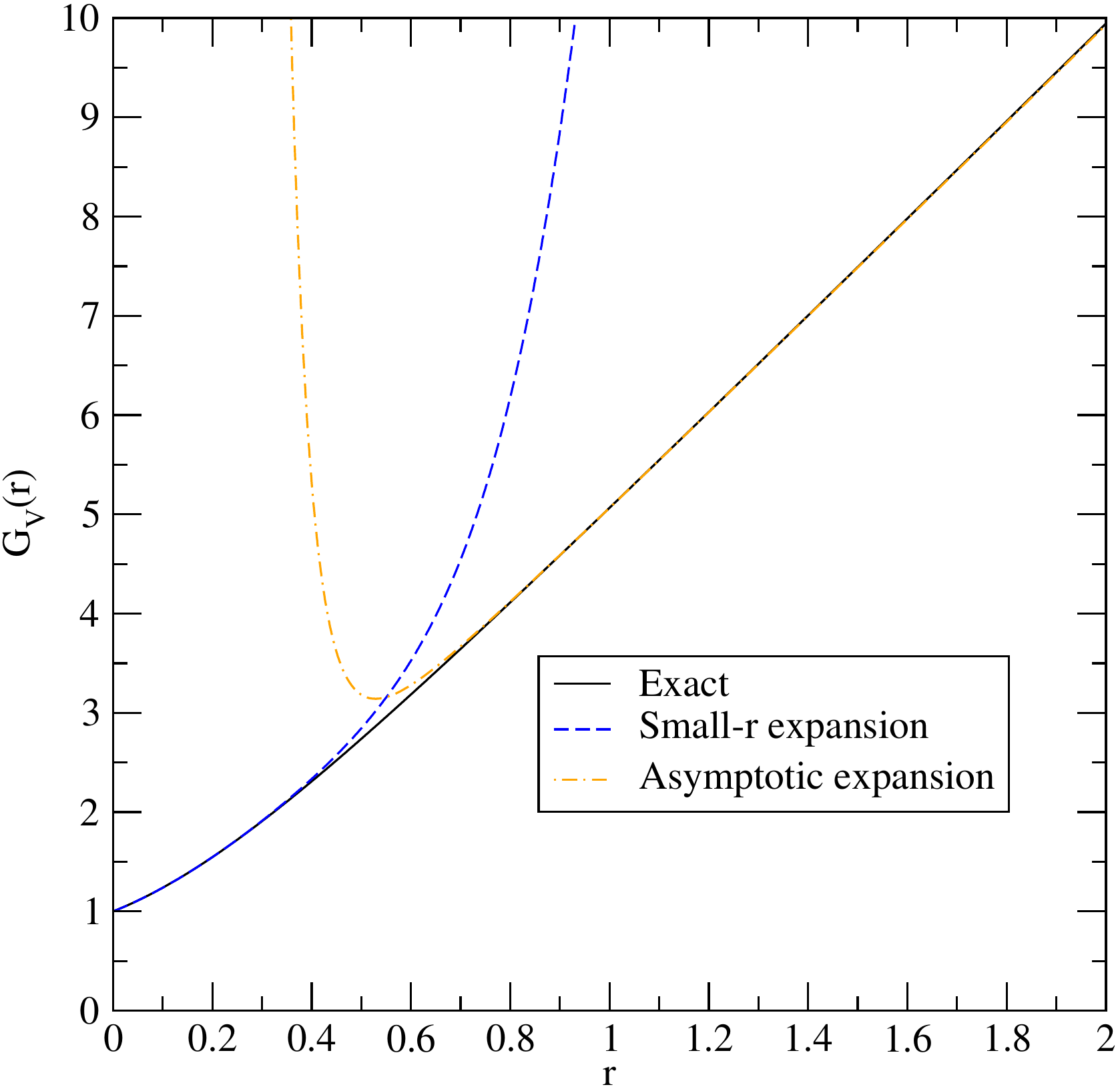}
\caption{Comparison of the exact form of $G_V$ for the $d=1$ determinantal point process with the
small- and large-$r$ expansions in \eqref{Gvsmalls} and \eqref{Gvlarges}.}\label{tone}
\end{figure}

\ASd{The asymptotic linear slope of $\pi^2/2$ is exactly as predicted by the leading-order term in the asymptotic analysis of \eqref{eq:PV}.  The results in \eqref{Gvsmalls} and \eqref{Gvlarges} have been checked by back-substitution into \eqref{PV} and by comparing the results to the exact form for $G_V$ via numerical calculation of the appropriate determinant.  Both expressions back-substitute to 0 for small- and large-$s$, respectively, and the results of the comparison are shown in Figure \ref{tone}.}

By looking at Figure \ref{tone}, one can see that the expansions are quite good for the ranges in $s$ where they are valid. Equations (\ref{PV}), (\ref{Gvsmalls}), and (\ref{Gvlarges}) constitute the solution to our problem. Although it is natural to ask if there is a corresponding nonlinear differential equation that characterizes $G_V$ in higher dimensions, we are not aware of any work in this direction, and this issue remains an open problem. 

\subsection{Two-dimensional processes}

There are a few examples of determinantal point processes in two dimensions. The seminal example is provided by the complex eigenvalues of random non-Hermitian matrices \cite{ginibre:440,forrester1998ert}. The kernel of such a determinantal point process is given by:
\begin{equation}\label{Ginibre}
H_N(z,w)=\left(\frac{1}{\pi}\right)\exp\left[-\frac{1}{2}(|z|^2+|w|^2)\right]\sum_{k=0}^{N-1}\frac{(z \overline{w})^k}{k!},
\end{equation}
where $N$ is the rank of the matrix and $z,w\in\mathbb{C}$. Incidentally, \eqref{Ginibre} can be related to the distribution of $N$ polarized electrons in a perpendicular magnetic field, filling the $N$ lowest Landau levels. In the limit $N\to \infty$ \eqref{Ginibre} becomes
\begin{equation}
\label{eq:Kgauss}
H(z,w)=\left(\frac{1}{\pi}\right)\exp\left[-\frac{1}{2}(|z|^2+|w|^2-2z\overline{w})\right],
\end{equation}
which is a homogeneous and isotropic process ($\rho=H(z,z)=\frac{1}{\pi}$) in $\mathbb{C}$. It is instructive to examine the pair correlation function, which after some algebra can be written as:
\begin{equation}
\label{eq:g2gauss}
g_2(z_1,z_2)=1-\exp\left(-|z_1-z_2|^2\right).
\end{equation}
From this expression one finds that the correlation between two points decays like a Gaussian with respect the distance separating the points. 
Letting $r = \lvert z_1 - z_2\rvert$, we may write the associated structure factor of the system as:
\begin{equation}\label{ginibreSk}
S(k) = 1-\exp\left[-\frac{k^2}{4}\right],
\end{equation}
which has the following small-$k$ behavior:
\begin{equation}\label{ginibreSksmallk}
S(k) \sim \frac{k^2}{4} + \Ord{k^4} \qquad (k\rightarrow 0).
\end{equation}
We see that the determinantal point process generated by the Ginibre ensemble is hyperuniform with an exponential scaling $\alpha = 2$ for small $k$,
corresponding to an endpoint of one of the ``allowed'' intervals for determinantal PPs; the large-$r$ behavior of the kernel $H(r)$ is exponential ($H(r) = \exp[-r]$).

Other ensembles of two-dimensional determinantal point processes can be found in simple systems. For example, the $n$ zeros of 
an analytic Gaussian random function $f(z)=\sum_{k=0}^n a_k z^k$ also form a determinantal point process on 
the open unit disk \cite{peres2000zig, leboeuf1999rmr}. The limiting kernel governing these zeros is called the Bargmann kernel:
\begin{equation}
\label{eq:BargK}
H(z_1,z_2)=\left(\frac{1}{\pi}\right)\frac{1}{(1-z_1 \overline{z_2})^2}
\end{equation}
and is inherently different from \eqref{eq:Kgauss}. 

\section{An algorithm for generating determinantal point processes}

We are able to write down an algorithm, which we call the HKPV algorithm, after \cite{hough2006dpa}, to generate determinantal point processes due to the geometric interpretation of the determinant in $\rho_N$ as the volume of the simplex built with the $N$ vectors $v_j=\{\lvert\phi^0_j\rangle\}_{1\leq j\leq N}$. In the original paper \cite{hough2006dpa} this algorithm is sketched and then proved to produce the correct distribution function $p_N$. The algorithm is extremely powerful and versatile and we believe it is important to provide as many details as possible about it and its implementantion (which has not been done before, to our knowledge). Therefore we dedicate the present section to provide a complete description of the HKPV algorithm and enough details (with some tricks) for its efficient implementation. 

Set $H_N\equiv H$, the kernel of the determinantal point process. Pick a point $\mathbf{\xi}_N$ distributed with probability: 
\begin{equation}
p_N(\mathbf{x})=H_N(\mathbf{x},\mathbf{x})/N.
\end{equation}
With this point build the new operator $A_{N-1}$, defined by:
\begin{equation}
A_{N-1}=H_N\ket{\mathbf{\xi}_N}\bra{\mathbf{\xi}_N}H_N.
\end{equation}
This operator has with probability 1 a single nonzero eigenvalue and $N-1$ null eigenvalues. 
When expressed as a matrix in the basis $\{\lvert\phi^0_n\rangle\}_{1\leq n\leq N}$, $A_{N-1}$ takes the form:
\begin{equation}
\left(A_{N-1}\right)_{i,j}=\overline{\phi^{0}_i(\mathbf{\xi}_N)}\phi^{0}_j(\mathbf{\xi}_N).
\end{equation}
Consider the $N-1$ null eigenvectors of $A_{N-1}$; we will denote them as $\left\{\ket{\phi^1_i}\right\}_{i=1}^{N-1}$
and call $\ket{\phi^{1}_N}$ the only eigenvector with a nonzero eigenvalue. 
The null eigenvectors can be found easily by means of a fast routine based on singular value decomposition (SVD), but we will see one that can proceed without it. 

Next, build the new operator $H_{N-1}$:
\begin{equation}
H_{N-1}=H_N\left(\sum_{n=1}^{N-1}\ket{\phi^1_n}\bra{\phi^1_n}\right)H_N.
\end{equation}
To simplify the computation, notice that by completeness of the basis $\left\{\ket{\phi^1_n}\right\}_{n=1}^N$ in the eigenspace of $H_N$: 
\begin{equation}
H_N\left(\sum_{n=1}^{N-1}\ket{\phi^1_n}\bra{\phi^1_n}+\ket{\phi^1_N}\bra{\phi^1_N}\right)H_N=H_{N},
\end{equation}
and since $\ket{\phi^1_N}$ is the only eigenvector orthogonal to the null space:
\begin{equation}
A_{N-1}=\text{tr}(A_{N-1})\ket{\phi^1_N}\bra{\phi^1_N}.
\end{equation}
From this equation we conclude:
\begin{equation}
H_{N-1}\equiv H_N\left(\sum_{n=1}^{N-1}\ket{\phi^1_n}\bra{\phi^1_n}\right) H_N=H_N\left(\mathbb{I}-\frac{1}{\text{tr}(A_{N-1})}A_{N-1}\right)H_N.
\end{equation}
Once $H_{N-1}$ is obtained, we repeat the procedure with $H_N\to H_{N-1}$, generating the point $\mathbf{\xi}_{N-1}$ from the probability distribution: 
\begin{equation}
p_{N-1}(\mathbf{x})=H_{N-1}(\mathbf{x},\mathbf{x})/(N-1)
\end{equation}
and the operators $A_{N-2},~H_{N-2}$.  As the number of iterations increases, we constantly reduce the rank of the operators by one: 
$\text{tr}(H_{N})=N$, $\text{tr}(H_{N-1})=N-1$, etc. 
Therefore, after we have placed the last point $\mathbf{\xi}_1$, we are left with the an operator of rank 0, and the algorithm stops.
Reference \cite{hough2006dpa} shows that the $N$-tuples $(\mathbf{\xi}_1,\dotsc,\mathbf{\xi}_N)$ are distributed according to the distribution (\ref{eq:probdef}).

The whole procedure requires $\Ord{N^2}$ steps for every realization, which is equal to the number of function evaluations necessary to create the matrices $A$. Therefore, the algorithm is computationally quite light. The only subroutine that requires some work is the extraction of the random points from the probability distributions $p_{n}(\mathbf{x})$. For $d = 1$ one can use a numerically-implemented inverse CDF technique \cite{devroye1986nur}, and the computational cost of this procedure is independent of $N$. For $d\geq 2$ if the distributions are not very peaked, a rejection algorithm is sufficient.  The rejection algorithm works by sampling points from a uniform distribution on the
domain.  A tolerance value near the maximum of the probability density of the point process is set, and the point is accepted if a uniform random number
chosen between 0 and the tolerance value is less than the probability density at that point.  Otherwise, the point is rejected, and the process repeats.   
Unfortunately, it is difficult to estimate the computational cost of this algorithm as a function of the number of particles $N$.

\subsection{Numerical results in one dimension}

We have implemented the algorithm described above to study a determinantal point process on the circle $x\in [0,2\pi]$, where $\phi_n(x)=\exp[inx]/\sqrt{2\pi}$ are the $N$ orthonormal functions with $n=0,\pm 1,\pm 2,\dotsc,\pm N/2$. This ensemble, as mentioned above, is equivalent to the one generated by the eigenvalues of unitary random matrices chosen according to the Haar measure. Eigenvalues of matrices from the CUE can be generated easily by means of a fast SVD algorithm \cite{mezzadri2006grm}; however, plenty of exact results exist. Therefore, we study this one-dimensional determinantal process as a test both for the performance of our algorithm and for the convergence of the results to the $N\to\infty$ limit.

We have implemented the algorithm in both Python and C++, noticing little difference in the speed of execution, and run it on a regular desktop computer. As mentioned above, the algorithm runs polynomially in $N$ with the sampling of the distribution $p_n$ limiting the computational speed. One point, however, which requires attention is the loss of precision of the computation. Due to the fixed precision of the computer calculations, the matrix $H_n$ ceases to have exactly integral trace, diminishing the reliability of the results. Typically, one observes deviations in the 5th decimal place after $\approx 50$ particles have been placed. We have devised an `error correction' procedure in which the numerical matrix $H_n$ is projected onto the closest Hermitian matrix $\tilde{H}_n$ which has eigenvalues 1 or 0 only. This projection corrects for a great part, but not all, of the error; however, the algorithm is slowed by this modification. The number of particles in each configuration can therefore be pushed to $N\approx 100$, regardless of the dimension.  We have been able to generate between 75000 and 100000 configurations of points
in each dimension.
In general, the error-correction procedure generates more reliable statistics for a given value of $N$ compared to the uncorrected algorithm, 
and we therefore expect that any residual error not captured by the 
matrix projection is minimal.   

\begin{figure}[!htbp]
\centering
\includegraphics[width=0.45\textwidth]{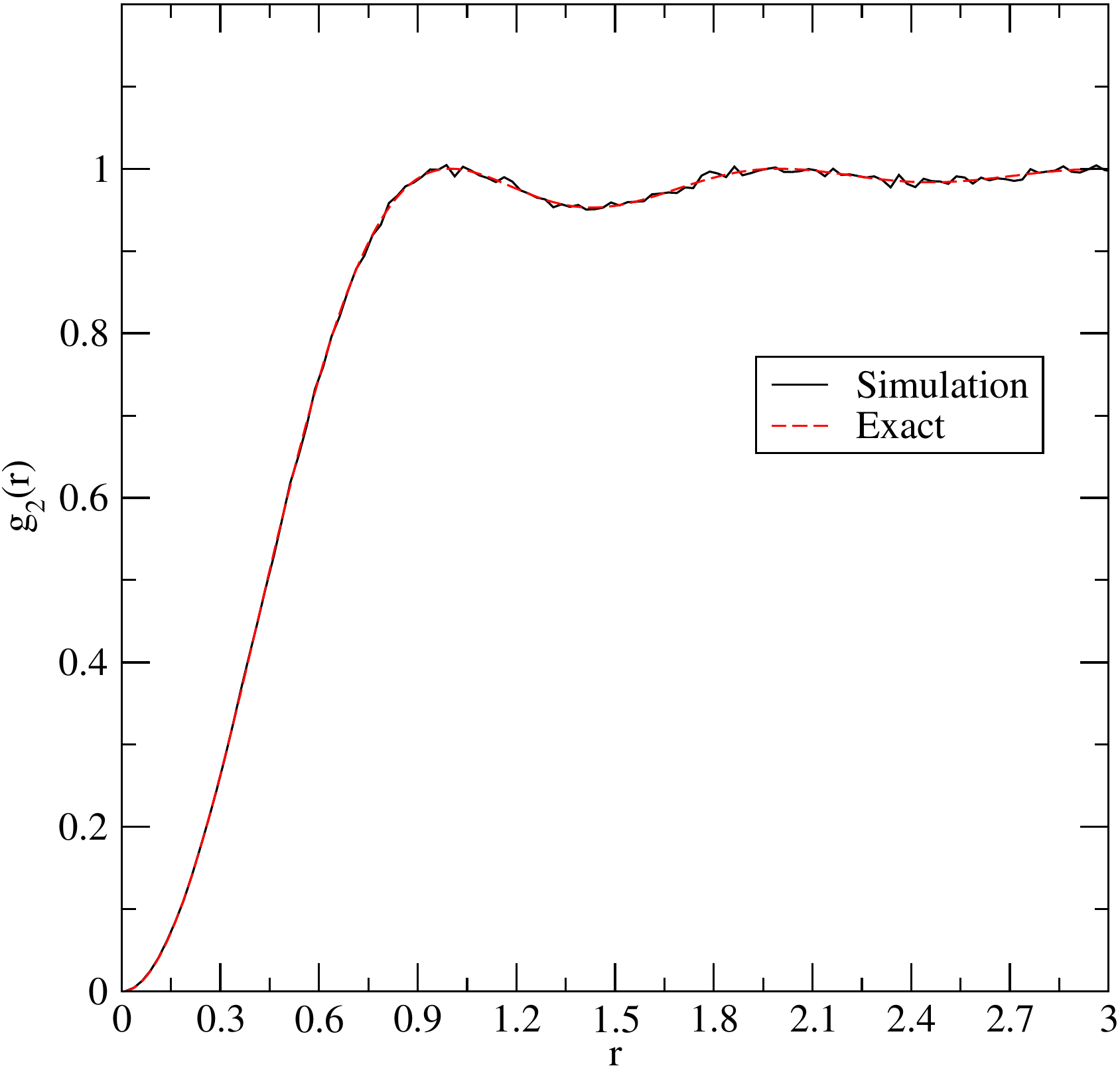}
\caption{Comparison of the exact expression \eqref{fermiong2} for $g_2(r)$ with the results from the HKPV Algorithm for $d = 1,~\rho = 1.$
The results from the simulation are obtained using 75000 configurations of 45 particles.}\label{1dg2}
\end{figure}

As a preliminary check for the HKPV Algorithm, 
we have calculated the pair correlation function $g_2$ and compared the results to the exact expression in \eqref{fermiong2}
below.
The comparison is quite favorable and suggests that the point configurations are being generated correctly by the algorithm.  We mention a few characteristics of $g_2$ 
which arise from the determinantal nature of the point process.  First, the system is strongly correlated for a significant range in $r$, and $g_2 (r) \rightarrow 0$ as 
$r \rightarrow 0$.  This \emph{correlation hole} \cite{Mc60, Sl51, BoCo74, Sc05} 
is indicative of a strong effective repulsion in the system, especially for small point separations.  In other words,
the points tend to remain relatively separated from each other as they are distributed through space.  Second, $g_2 \leq 1$ for all $r$, meaning that it is always
\emph{negatively correlated}; again, this quality is indicative of repulsive point processes, which are characterized by a reduction of the probability density near
each of the coordination shells in the system.  
We show in a separate paper \cite{ToScZa08} that at  \emph{fixed number density},
$g_2$ approaches an effective pair correlation function 
$g_2^*(r) = \Theta(r-D)$ as $d \rightarrow \infty$, suggesting that the points achieve an increasingly strong effective hard core $D~(\sim\sqrt{d})$ 
as the dimension of the system 
increases.  At \emph{fixed mean nearest-neighbor separation} this observation implies that 
$g_2(r) \rightarrow \tilde{g}_2^*(r) = 1$ for all $r>0$ as $d \rightarrow \infty$  (as $g_2(0)=0$ for any $d$), implying that
the points become completely uncorrelated in this limit.  We will show momentarily that the latter limit is difficult to interpret due to the
dimensional dependence of the density $\rho$.    

\ASd{\begin{figure}[!htbp]
\centering
\includegraphics[width=0.45\textwidth]{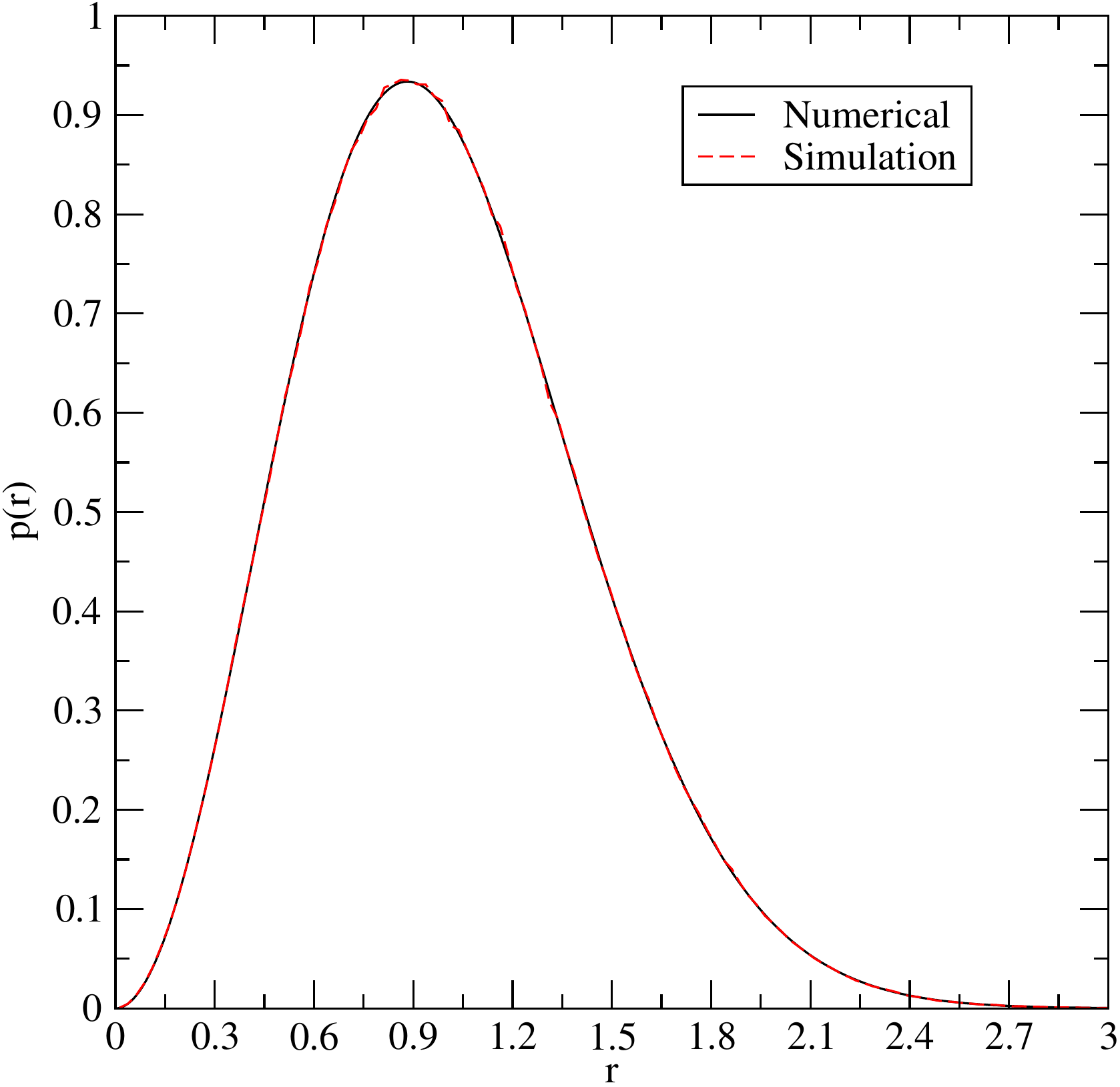}
\caption{Numerical and simulation results for the gap distribution function $p(r)$ for $d = 1,~\rho = 1$. Numerical results are 
obtained using \eqref{eq:PD0M}, and simulations were performed with the HKPV Algorithm under the same conditions as in Figure \ref{1dg2}.}\label{1dp}
\end{figure}}

\begin{figure}[!htbp]
\centering
\includegraphics[width=0.35\textwidth]{fermion1dpnumsimrho1}
\includegraphics[width=0.36\textwidth]{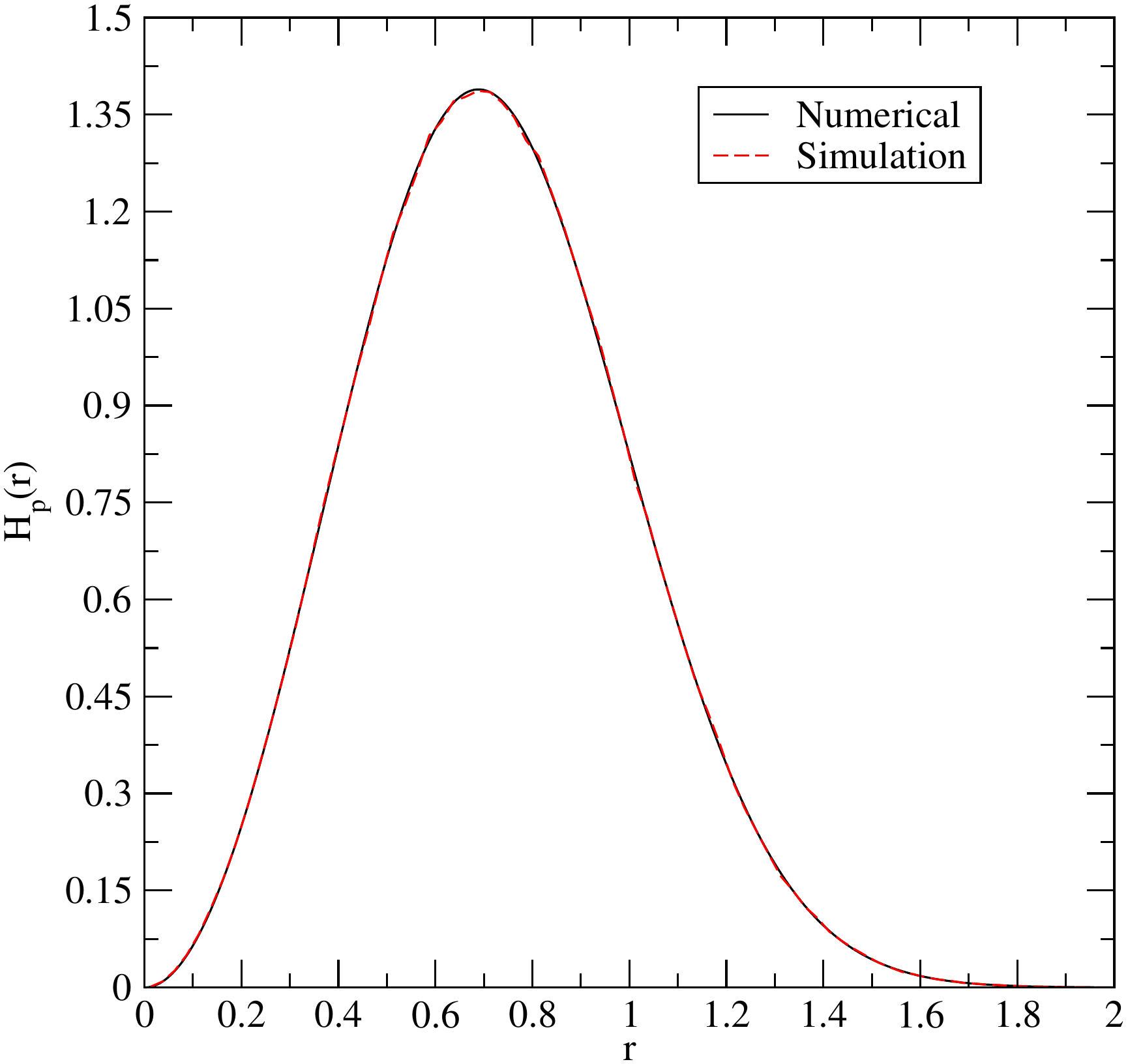}
\includegraphics[width=0.36\textwidth]{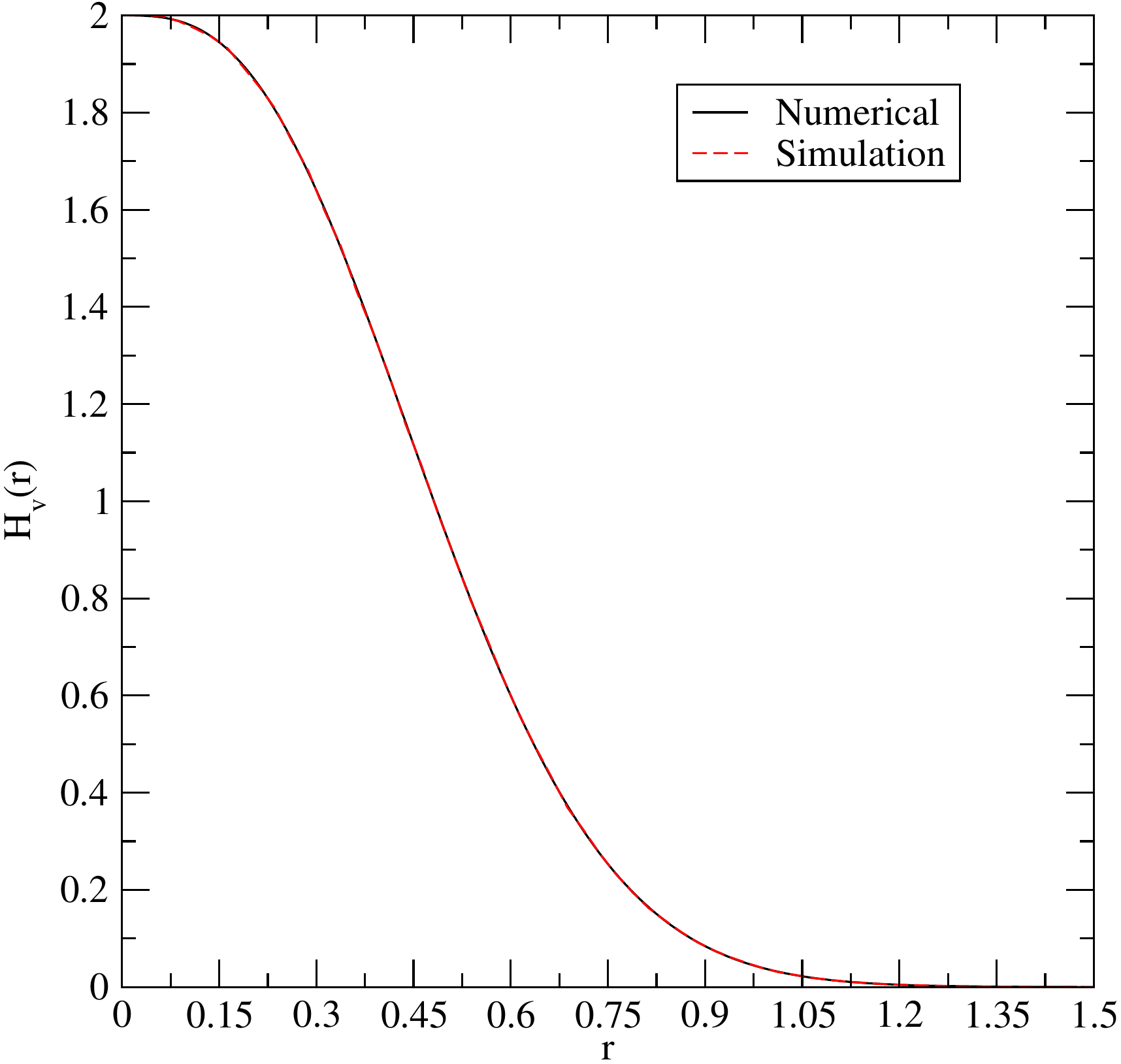}
\caption{Comparison of numerical and simulation results with $d = 1,~\rho = 1$ for \emph{left:} the gap distribution function $p(r)$, \emph{center:} $H_P(r)$ and
\emph{right:} $H_V(r)$.}\label{fig:pHpHv1d}
\end{figure}

Figure \ref{fig:pHpHv1d} presents the results for the gap distribution function $p(r)$ using both the HKPV Algorithm and a numerical calculation based on the 
determinant in \eqref{eq:PD0M}.  As with the calculation of $g_2$, the comparison between the numerical results and the simulation is favorable.  This 
curve, as expected, has the same form as the one reported in the random matrix literature \cite{mehta2004rm} and scales with $r^2$ as $r\rightarrow 0$.
We stress, however, that this function represents the distribution of gaps between points on the line and does not discriminate
between gaps to the left and to the right of a point.  The random matrix literature oftentimes 
describes this quantity as a ``nearest-neighbor'' distribution, which it 
is \emph{not}.  As mentioned in the discussion following \eqref{Hvdef} and \eqref{Hpdef}, the void and particle nearest-neighbor distribution functions are given
by the functions $H_V$ and $H_P$, respectively, and require that distance measurements be made \emph{both} to the left and to the right of a point; the numerical and
simulation results for these functions are also given in Figure \ref{fig:pHpHv1d}.        

The function $H_V$ is clearly different from $p$. $H_P$ has a similar shape to the gap distribution function; however, $H_P$ peaks more sharply around
$r \approx 0.725$ while $p$ has a less intense peak near $r \approx 1$.  This observation is justified from a numerical standpoint since point separation measurements are
made in both directions from a given reference point with only the \emph{minimum} separation contributing to the final histogram of $H_P$.  In contrast, \emph{every} gap 
in the point process is used for constructing the histogram of $p$.  As a result, we expect the first moment of $H_P$ to be less than that of $p$, and this result
is exactly what we observe in Figure \ref{fig:pHpHv1d}.  

\begin{figure}[!htbp]
\centering
\includegraphics[width=0.45\textwidth]{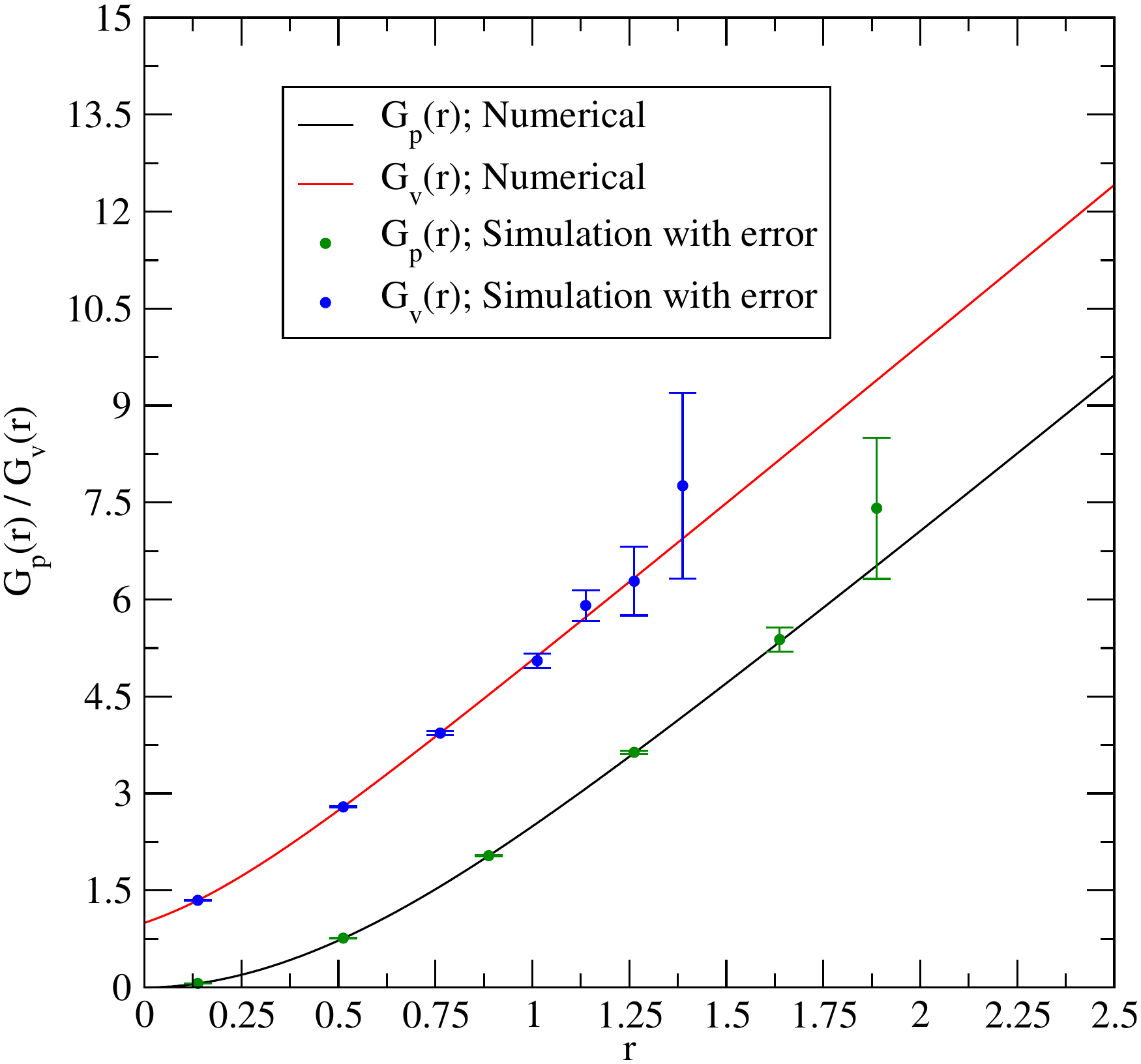}
\caption{Numerical results using \eqref{eq:PD0M} for $G_P$ and $G_V$ with $d = 1,~\rho = 1$.  Also included are representative simulation results and estimated errors
from the HKPV Algorithm under the same conditions as in Figure \ref{1dg2}.}\label{1dGpGv}
\end{figure}

The form of $H_V$ may at first seem confusing in the context of our discussion above concerning the inherent repulsion of the determinantal point process.  Unlike $H_P$ and 
$p$, the void nearest-neighbor function $H_V$ has a nonzero value at the origin and is monotonically decreasing with respect to $r$.  To understand this behavior, it is 
useful to examine the behavior of the corresponding $G_V$ and $G_P$ functions, which are plotted in Figure \ref{1dGpGv}.
We recall from \eqref{Gvdef} and \eqref{Gpdef} that $G_V$ and $G_P$ are related to
conditional probabilities which describe, given a region of radius $r$ empty of points (other than at the center for $G_P$), the probability
of finding the nearest-neighbor point in a spherical shell of volume $s(r)dr$, where $s(r)$ is the surface area of a $d$-dimensional sphere of radius $r$. Of particular 
relevance to the behavior of $H_V$ is the fact that $G_V(0) = 1$ and $s(0) = 2$ for $d = 1$. Therefore, the dominant factor controlling the small-$r$ behavior of $H_V$ is the spherical surface area $s(r)$ \cite{ToScZa08}. 
Since $s(0)$ is nonzero for $d = 1$, it follows from \eqref{Gvdef} that $H_V(0)$ is nonzero in contrast to $H_P(0)$.  

The behavior of both $G_P$ and $G_V$ is of particular interest in this paper.  We conjecture that both functions are linear for sufficiently large $r$ in 
\emph{any} dimension.  We show elsewhere \cite{ToScZa08} that as $r\rightarrow 0$, $G_P(r)\sim \xi(d)r^2+\Ord{r^4}$ and $G_V(r) \sim 1+\Ord{r^d}$,
where $\xi(d)$ is a dimensionally-dependent constant (for $d = 1$ this is evident in Figure \ref{1dGpGv}). Additionally, we believe that $G_V$ and $G_P$ obtain the same slope in the large-$r$ limit, and we will provide further commentary on this notion momentarily (see Fig.\ref{Gtilde}).
It is clear from Figure \ref{1dGpGv} that the results from the simulations are in agreement with the numerical results for a wide but limited
range of $r$ and they begin to deviate respectably for $r$ sufficiently large. This is due to the fast decay of both $H_{P/V}$ and $E_{P/V}$ to zero, therefore giving very small statistics (and a large degree of uncertainty) at these values of $r$.  This said, the numerical results are clear and provide strong support for our claims above.        

\subsection{Fermi sphere determinantal point process for $d \geq 2$}

\subsubsection{Definition of the Fermi sphere point process}

Here we study the determinantal point process of free fermions on a torus, filling a Fermi sphere. A detailed description of this process in any dimension may be found in an accompanying paper \cite{ToScZa08}.  We consider this example because it is the straightforward generalization of the one-dimensional CUE process described above. However, 
sampling of this ensemble cannot be accomplished with methods other than the algorithm introduced above; this limitation is in contrast to the two examples from Section III.C, where the ensemble may be generated from zeros of appropriate random complex functions. Nevertheless, it is difficult to construct another procedure that can be generalized to higher dimensions since zeros of complex functions and random matrices are naturally constrained to $d\leq 2$. 

We consider the determinantal point process obtained by ``filling the Fermi sphere" in a $d$-dimensional torus, i.e., $\mathbf{x}\in [0,2\pi]^d$; our choice of the
box size is for convenience and without loss of generality. We therefore consider all functions 
of the form:
\begin{equation}
\phi_{\mathbf{n}}=\left(\frac{1}{2\pi}\right)^{d/2}\exp\left[i(\mathbf{n}, \mathbf{x})\right]
\end{equation}
with
\begin{equation}
\label{eq:nmKF}
\lVert\mathbf{n}\rVert^2\leq \kappa_F^2(N),
\end{equation}
where $\kappa_F^2(N)$ is implicitly defined by the total number of states contained in the reciprocal-space sphere. This process is translationally-invariant for any $N$, both finite or infinite, and isotropic in the limit $N\to \infty$; it possesses the symmetry group of the boundary of the set (\ref{eq:nmKF}), a dihedral group which approximates $SO(d)$ very well for $N$ sufficiently large. The pair correlation function can be easily calculated for any $N<\infty$, and it is well-defined in the thermodynamic limit: 
\begin{equation}
g_2(x)=1-\frac{1}{N^2}\sum_{\mathbf{n}}\sum_{\mathbf{n}^{\prime}}\exp[i(\mathbf{n}-\mathbf{n}^{\prime}, \mathbf{x})],
\end{equation}
where $\mathbf{n}$ and $\mathbf{x}$ are $d$-dimensional real vectors, and the sums extend over the set (\ref{eq:nmKF}), which contains $N$ points. In the limit $N\to \infty $ the sums become integrals over a sphere of radius $k_F=2\sqrt{\pi}\left[\rho\Gamma(1+d/2)\right]^{1/d}$, where $\rho = N/(2\pi)^d$ is the number density.
The resulting pair correlation function is given by:
\begin{equation}\label{fermiong2}
g_2(r)=1-\left(\frac{2^d[\Gamma(1+d/2)]^2}{(k_F r)^d}\right)[J_{d/2}(k_F r)]^2,
\end{equation}
where $J_{d/2}$ is the Bessel function of order $d/2$ (cf.\ \cite{ToScZa08}). This pair correlation function is clearly different from (\ref{eq:g2gauss}) for $d = 2$; the two are therefore not equivalent, even in the thermodynamic limit. 
One can also find the limiting kernel:
\begin{equation}
H(\mathbf{x},\mathbf{y})=\left[\frac{(2^{d/2})\Gamma(1+d/2)}{(k_F \lVert\mathbf{x}-\mathbf{y}\rVert)^{d/2}}\right]J_{d/2}(k_F \lVert\mathbf{x}-\mathbf{y}\rVert),
\end{equation}
which is also different from (\ref{eq:Kgauss}) and (\ref{eq:BargK}) for $d = 2$.\footnote{Different fillings of the spectrum give raise to different families of determinantal point processes. In \cite{ToScZa08} we present another example of determinantal point process where states with momenta $k$ where $0\leq k\leq k_1$ or $k_2\leq k \leq k_3$ are filled. We called these Fermi-shell point processes.}

\begin{figure}[!htbp]
\begin{center}
\includegraphics[width=0.3\textwidth]{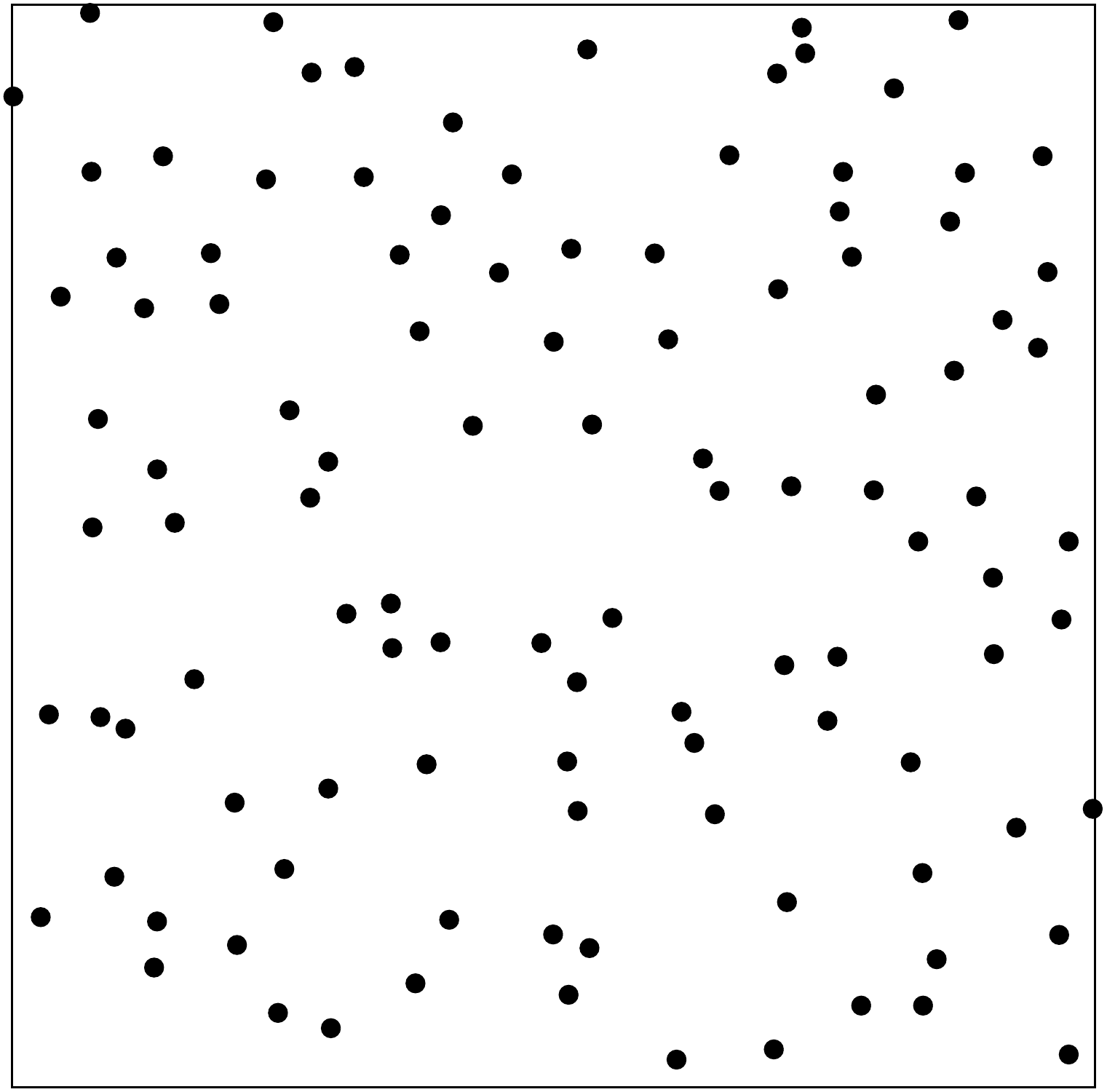}\hspace{0.5cm}
\includegraphics[width=0.3\textwidth]{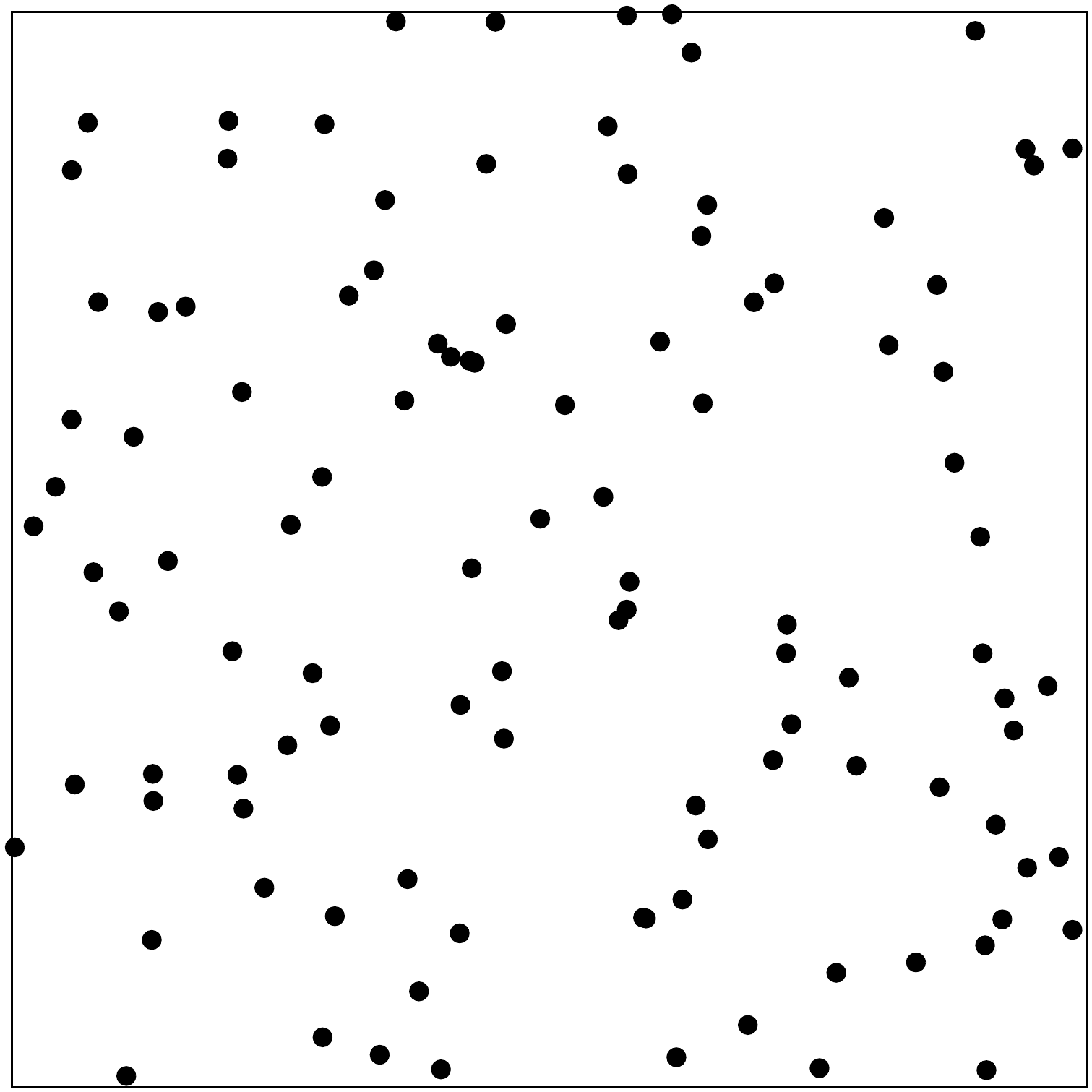}\\
\vspace{0.5cm}
\includegraphics[width=0.3\textwidth]{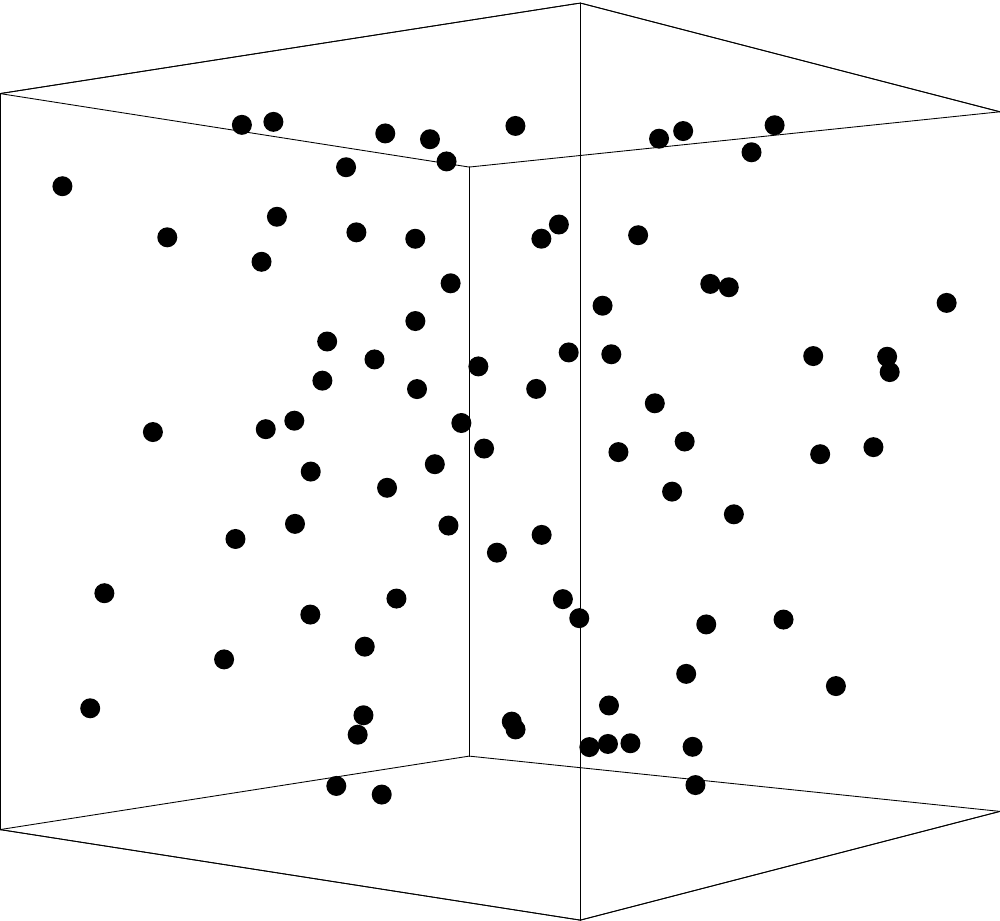}\hspace{0.02\textwidth}
\includegraphics[width=0.3\textwidth]{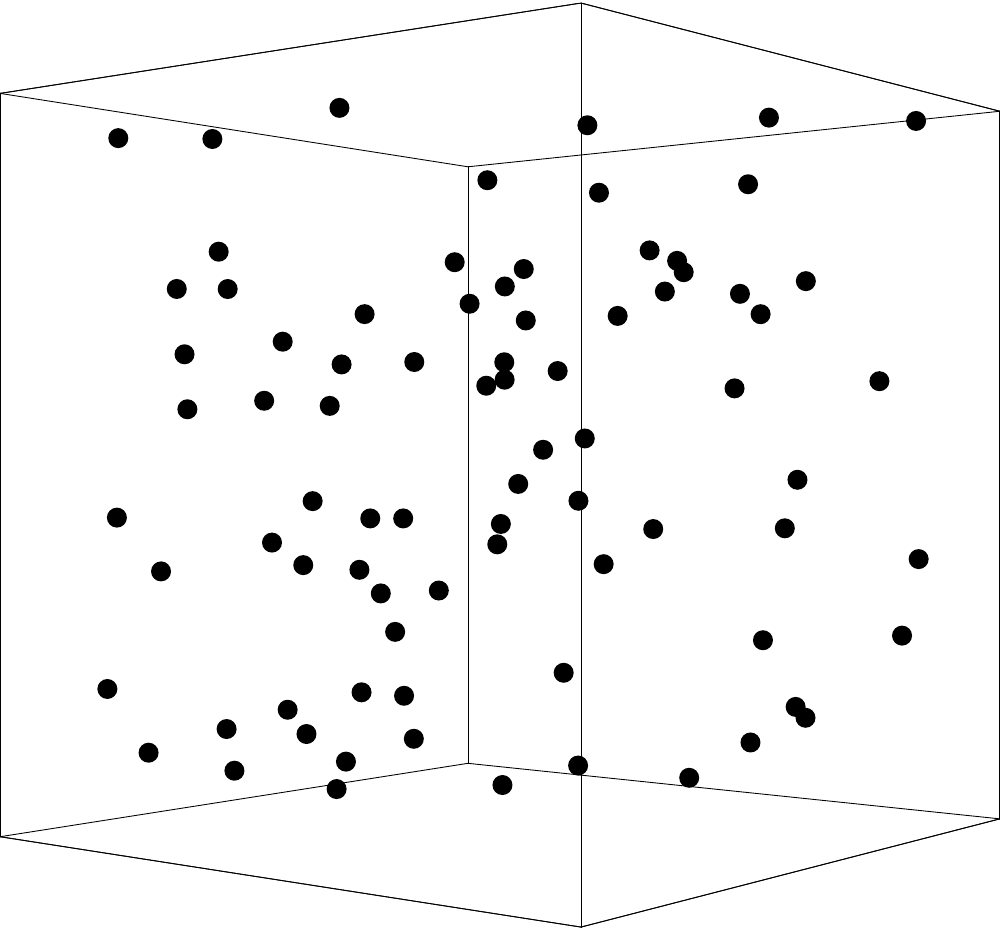}
\end{center}
\caption{\emph{Upper:} A $d = 2$ configuration of $N=109$ points distributed according to \emph{left:} a Fermi sphere determinantal point process and 
\emph{right:} a Poisson point process. \emph{Lower:}  A $d = 3$ configuration of $N=81$ points distributed according to \emph{left:} the Fermi sphere determinantal point process and \emph{right:}  a Poisson point process. All configurations have $\rho=1$.}

\label{fermpoi2d}
\end{figure}

Figure \ref{fermpoi2d} shows configurations of points generated for the $d = 2$ and $d = 3$ Fermi sphere point process alongside 
corresponding configurations for the Poisson point process in these dimensions.  The repulsive nature of the determinantal point process is immediately apparent
from these figures; note especially that the Fermi sphere point process discourages clustering of the points in space.  In contrast, clustering is not prohibited in the
Poisson point process, and small two- and three-particle clusters are easily identified.  Of particular interest is that the Fermi sphere point process  
distributes the points more evenly through space due to the effective repulsion in the system. This characteristic reflects the hyperuniformity of the point process
\cite{ToSt03}, and we will have more to say about this property momentarily.  

\ASd{
\begin{figure}[!htbp]
\centering
\caption{\emph{Left:}  A $d = 3$ configuration of the Fermi sphere determinantal point process, generated with the 
HKPV algorithm using $N = 81$ points and $\rho = 1$.\\  
\emph{Right:}  A $d = 3$ configuration of a Poisson point process with $N = 81$ points and $\rho = 1$.}\label{fermpoi3d}
\end{figure}}

\begin{figure}[!htbp]
\centering
\includegraphics[width=0.45\textwidth]{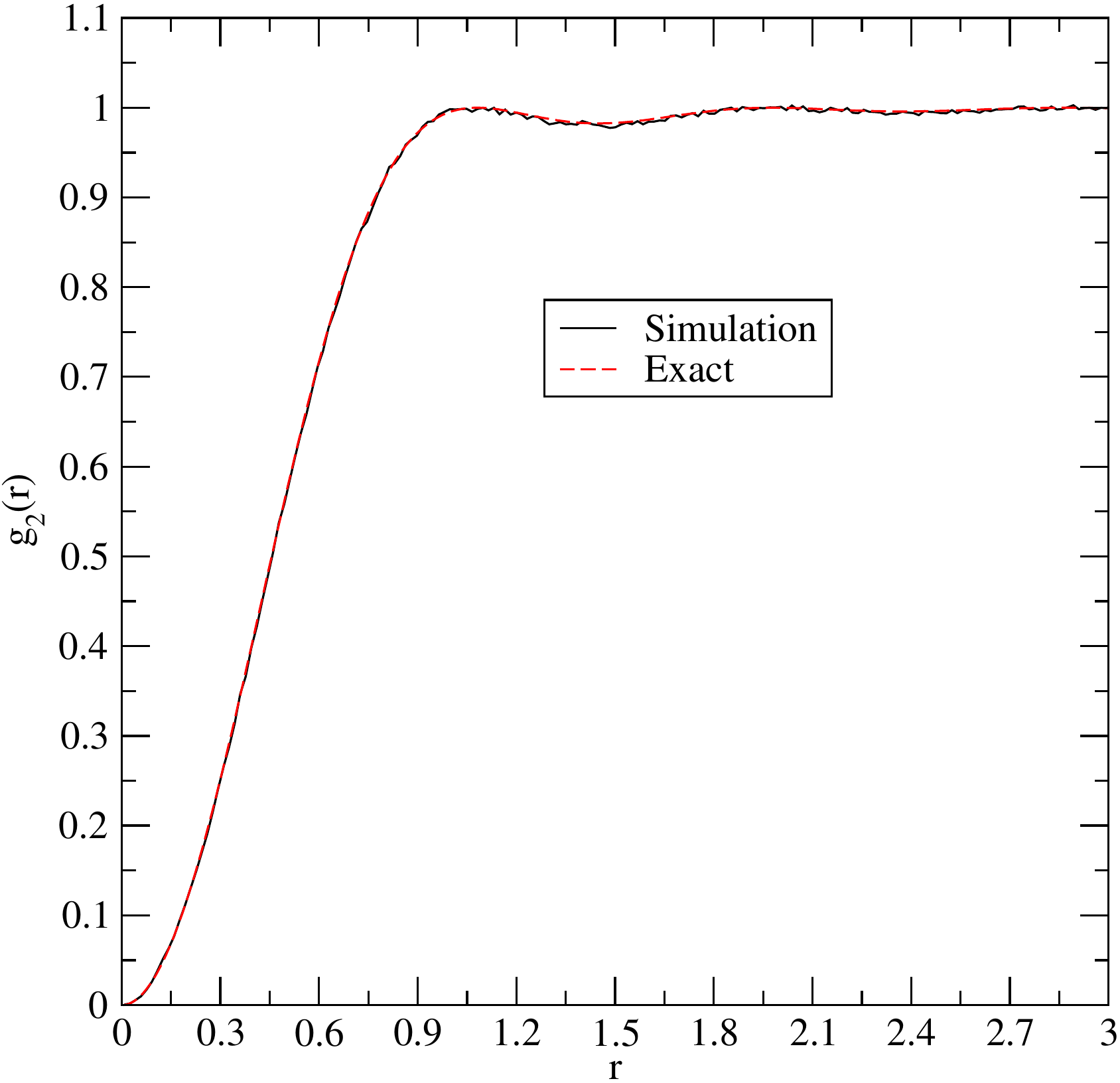}
\caption{Comparison of the exact expression \eqref{fermiong2} for $g_2(r)$ with the results from the HKPV Algorithm for $d = 2,~\rho = 1.$
The results from the simulation are obtained using 75000 configurations of 45 particles.}\label{2dg2}
\end{figure}

\subsubsection{Calculation of $g_2$ and nearest-neighbor functions for $d\geq 2$}

Figure \ref{2dg2} shows the numerical and simulation results for the pair correlation function $g_2$ with $d = 2$; a comparison of the results provides
strong evidence that the HKPV Algorithm correctly generates configurations of points for the Fermi sphere point process even in higher dimensions.
Note that the $d = 2$ correlations are significantly diminshed with respect to the form of $g_2$ for $d = 1$; this behavior is in accordance with 
a type of decorrelation principle \cite{ToSt06, ScStTo08} for the system.  Namely, we expect that as the dimension of the system increases,
unconstrained correlations in the system diminish.  We also remark that all higher-order correlation functions $g_n$ can be written in terms of the pair correlation 
function $g_2$ for \emph{any} determinantal point process.  
We prove this claim in an accompanying paper \cite{ToScZa08}.
It is therefore clear that the HKPV Algorithm is a powerful method by which one can study determinantal point processes in higher dimensions.

\begin{figure}[!htbp]
\centering
\includegraphics[width=0.4\textwidth]{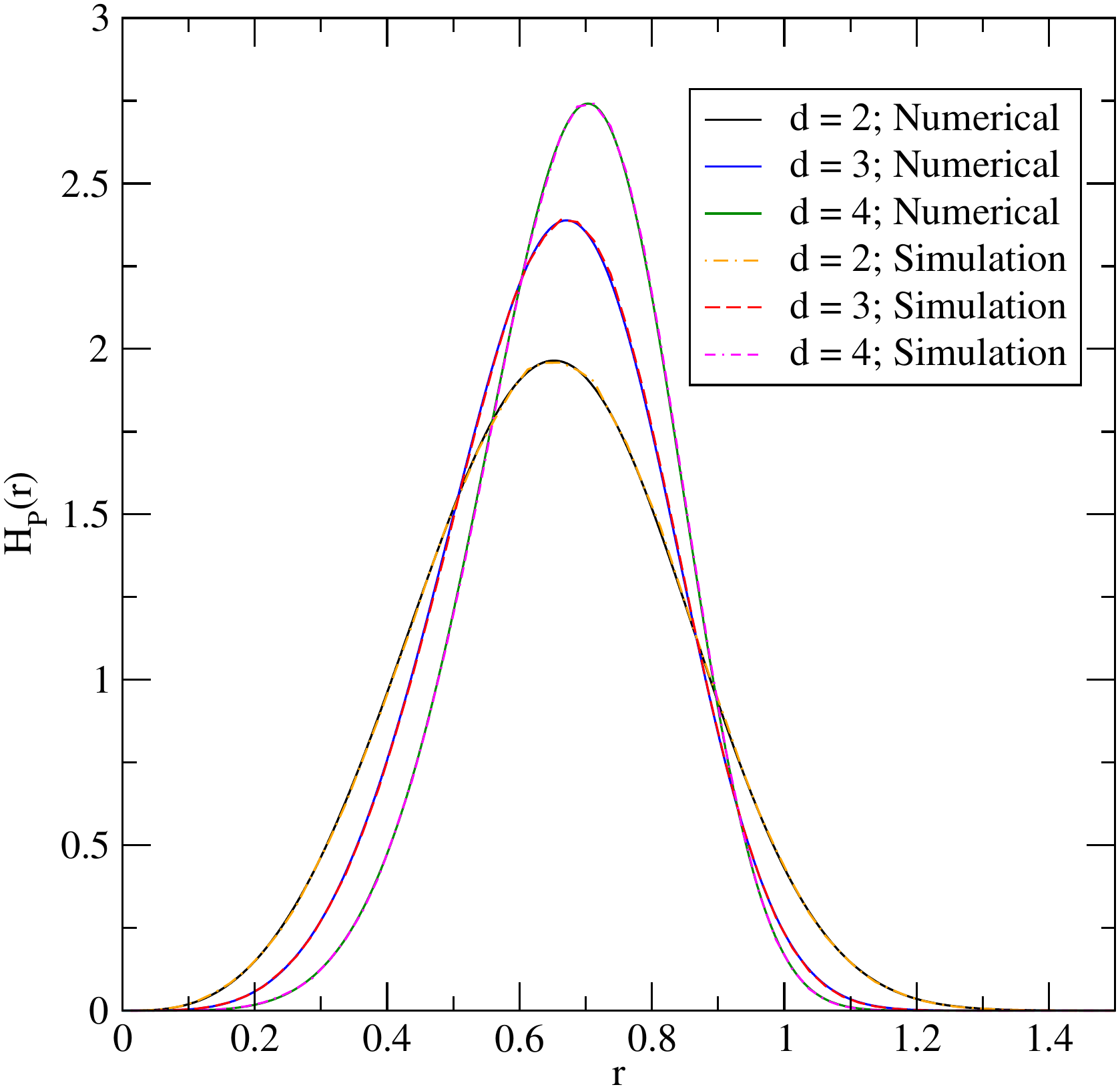}\hspace{0.5cm}
\includegraphics[width=0.4\textwidth]{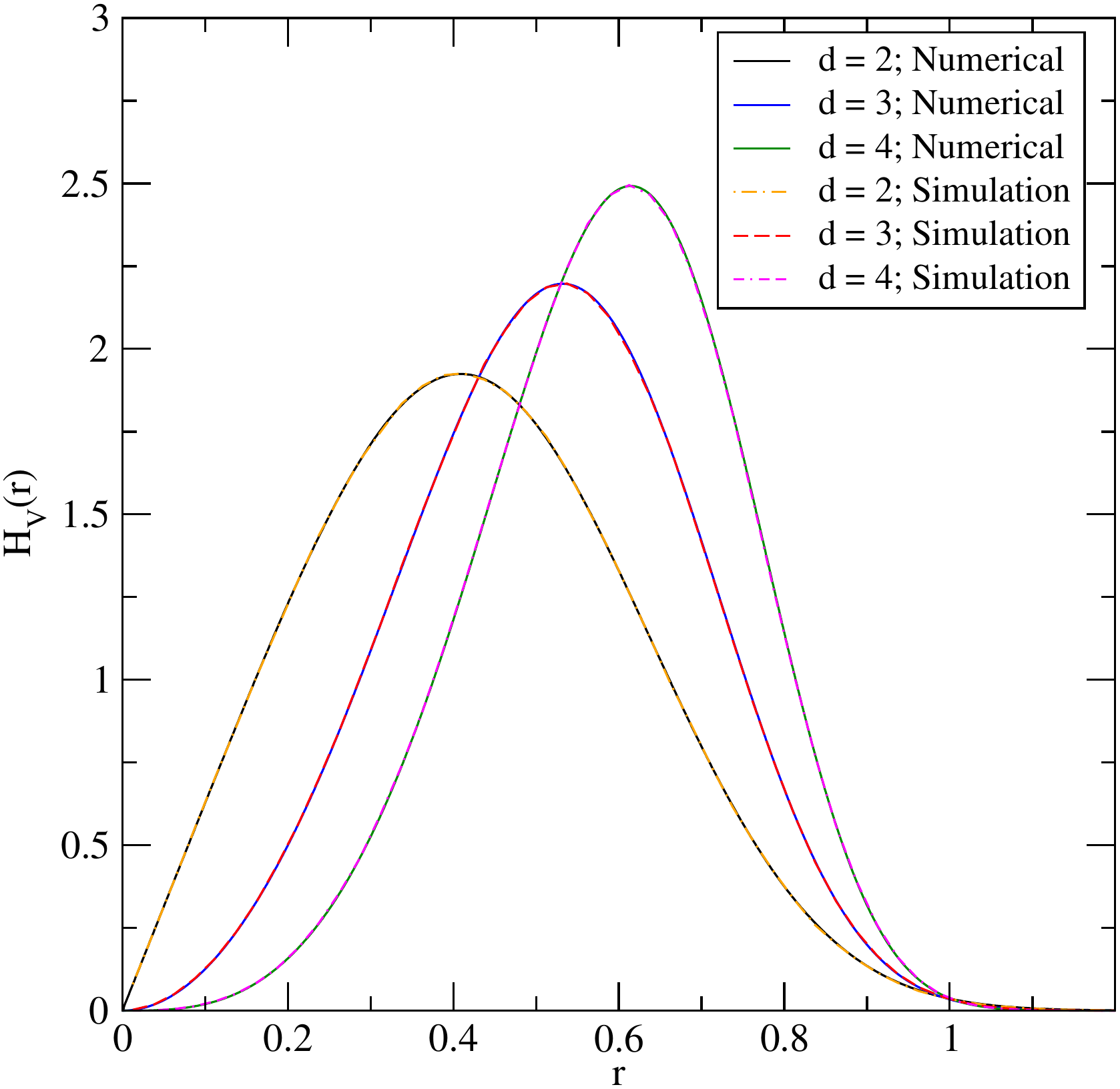}
\caption{Comparison of numerical and simulation results for \emph{left:} $H_P$ with $d = 2, 3$, and $4$ at number density $\rho = 1$ and \emph{right:} for $H_V$ with $d = 2, 3,$ and $4$ at number density $\rho = 1$.}\label{2d3d4dHpHv}
\end{figure}

Figure \ref{2d3d4dHpHv} contain results for the nearest-neighbor particle and void density functions $H_P$ and $H_V$ for $d = 2, 3,$ and $4$.  In all cases 
the numerical results coincide with the simulation results.  We do note that for $d = 3$ and $d = 4$ we have implemented the error-correction procedure described in Section
IV.A to increase the reliability of the simulation results as well as the particle numbers.  As mentioned above, running the algorithm without error correction 
generally results in a loss of precision in the trace of the kernel matrix $H$ during computation; the error introduced by this loss of precision as measured by 
deviation from the ``exact'' numerical results increases with respect to increasing particle number, and we notice that the errors are more acute
for $d = 3$ and $d = 4$.  Although some error still remains in the results even after projecting the matrix $H$ onto the nearest Hermitian projection matrix, the
results in these figures leave us confident that the computations are reliable.  

\begin{figure}[!htbp]
\centering
\includegraphics[width=0.36\textwidth]{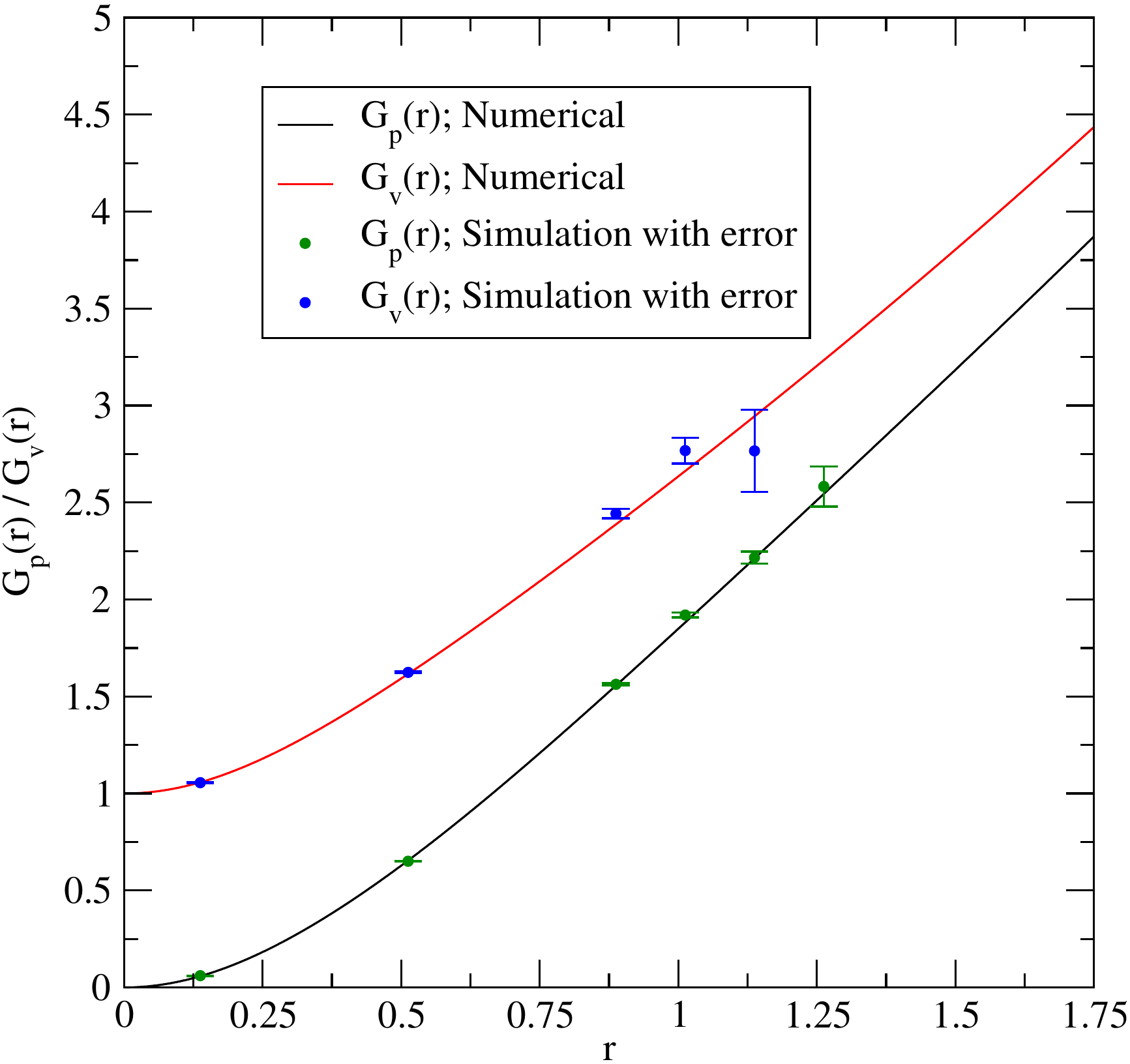}
\includegraphics[width=0.36\textwidth]{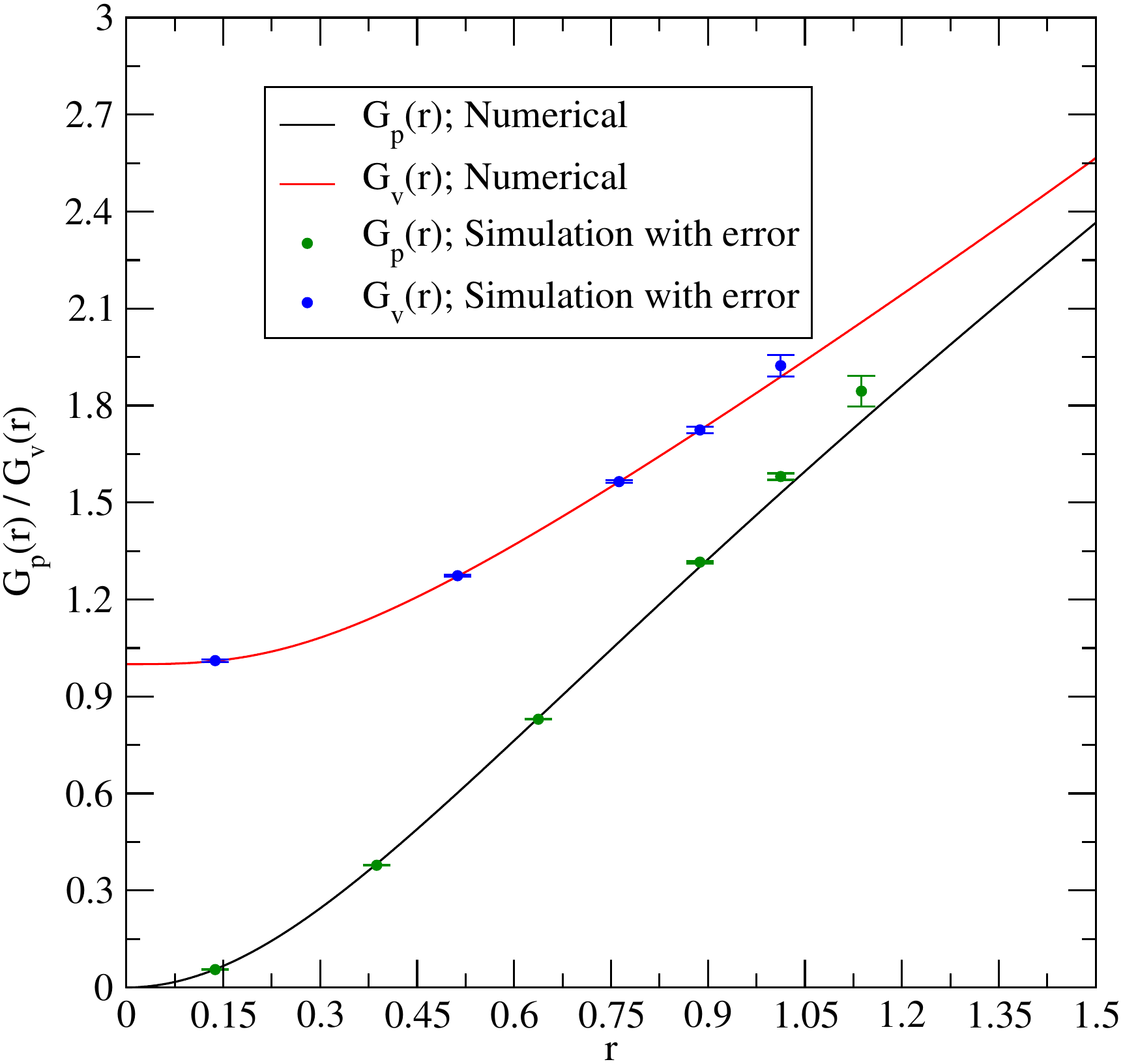}
\includegraphics[width=0.36\textwidth]{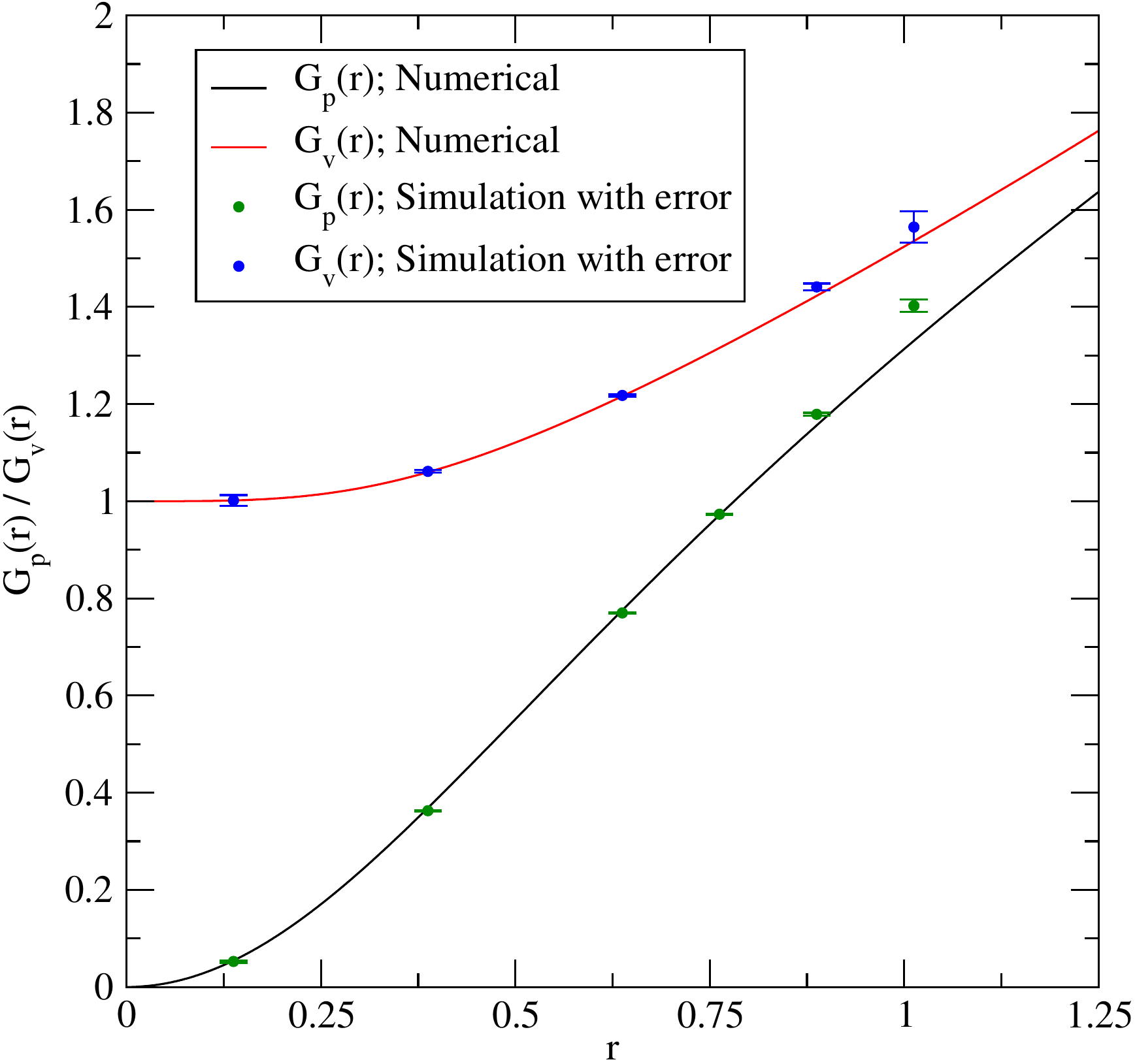}
\caption{Numerical results using \eqref{eq:PD0M} for $G_P$ and $G_V$ with (from left to right) $d = 2,3,4$ and for all $\rho=1$.  Also included are representative simulation results and estimated errors
from the HKPV Algorithm.}
\label{2dGpGv}
\end{figure}

In contrast to the $d = 1$ process, $H_V$ for $d = 2,3,4$ approaches $0$ as $r\rightarrow 0$; for $d = 3$ and $4$, $H_V$ and $H_P$ in fact possess very similar overall
shapes.  The small-$r$ behavior of $H_V$ in these cases is due to the behavior of $s(r)$ for $d \geq 2$; namely, $s(r) \sim r^{d-1}$ for all $d$, and for $d \geq 2$ we
observe that $s(0) = 0$ as opposed to the $d = 1$ case, where $s(0) = 2$.  We have already shown with generality that $G_V(r) \rightarrow 1$ as $r\rightarrow 0$,
a result which may be observed in Figures \ref{2dGpGv}.  One can see from these figures that $G_V(r) \rightarrow 1$ as $r\rightarrow 0$ in each 
dimension, reinforcing the dominance of $s(r)$ in the small-$r$ behavior of $H_V(r)$.  

For $d = 2$, the shape of $H_V$ resembles the corresponding curve for a Poisson point process; nevertheless, these two processes are inherently \emph{different}.
We may easily see the deviation between the two processes by noticing that
$H_P$ and $H_V$ do not coincide for any dimension and that $G_P$ and $G_V$ both increase linearly for sufficiently large $r$.  The latter observation
implies that $H_V(r) \nsim s(r) E_V(r)$ for large $r$, which is the case for the Poisson point process.  However, we
show elsewhere \cite{ToScZa08} that $E_V$ for the Fermi sphere point process in dimension $d$ (finite) 
behaves similar to the corresponding function for a Poisson point process
except in dimension $d+1$.  Further justification for this claim is also developed later is this paper.      

With regard to $H_P$, we remark that in each dimension $H_P(0) = 0$ in agreement with the repulsive nature of the point process.  However, 
it is worthwhile to note that, in light of the connection to noninteracting fermions described above, we can associate this repulsion with a type of
Pauli exclusion principle, which for noninteracting fermions is purely quantum mechanical in nature and arises solely from the constraint of 
antisymmetry of the $N$-particle wavefunction.  The determinantal form of the wavefunction is the manifestation of this antisymmetry in any dimension,
thereby providing some physical insight into the strong small-$r$ correlations for this determinantal point process.  
We stress that in the case of noninteracting fermions
the repulsion does not arise from any true interaction among the particles and is purely a consequence of the aforementioned antisymmetry.     

\begin{figure}[!htbp]
\centering
\includegraphics[width=0.365\textwidth]{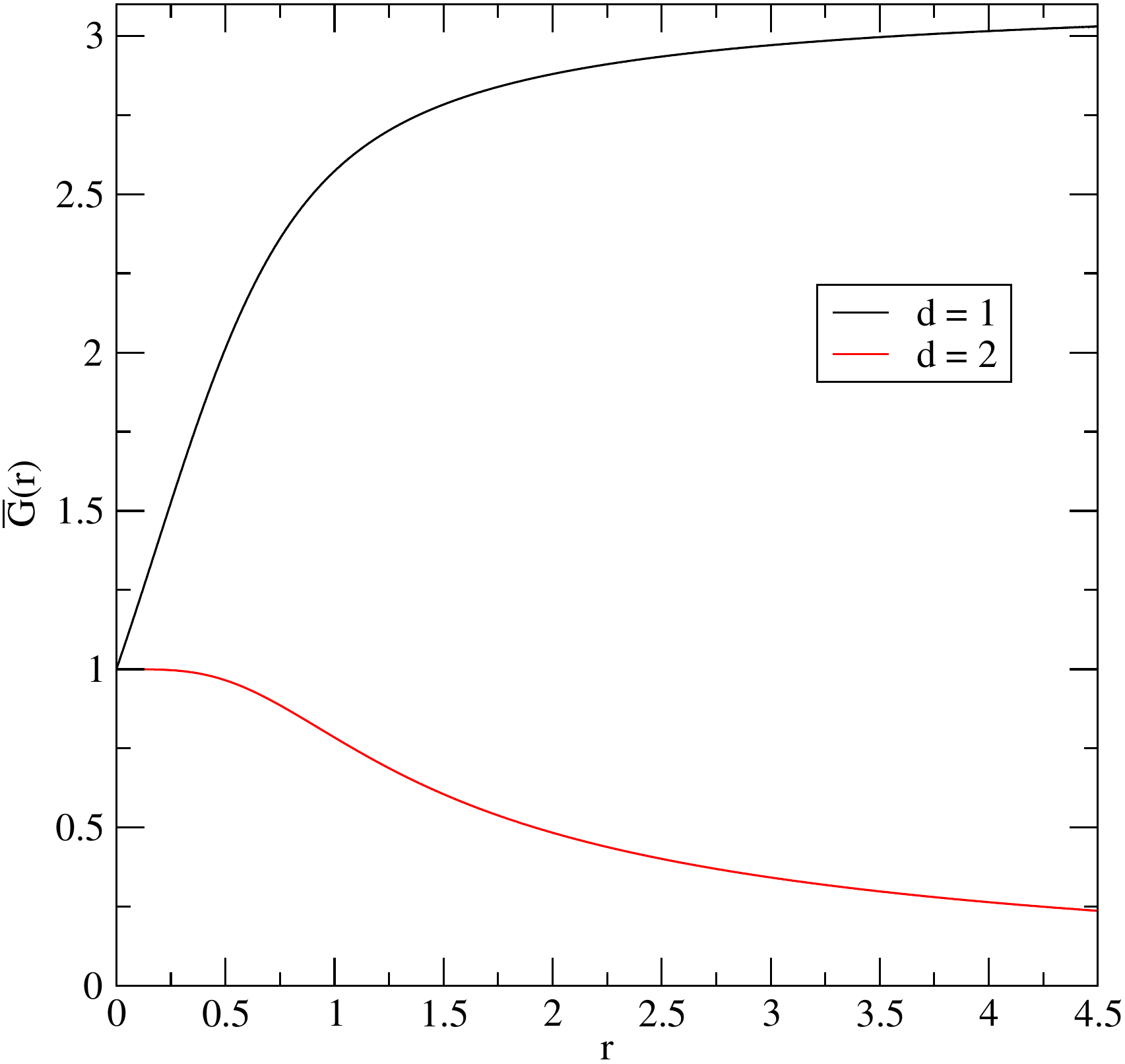}
\includegraphics[width=0.36\textwidth]{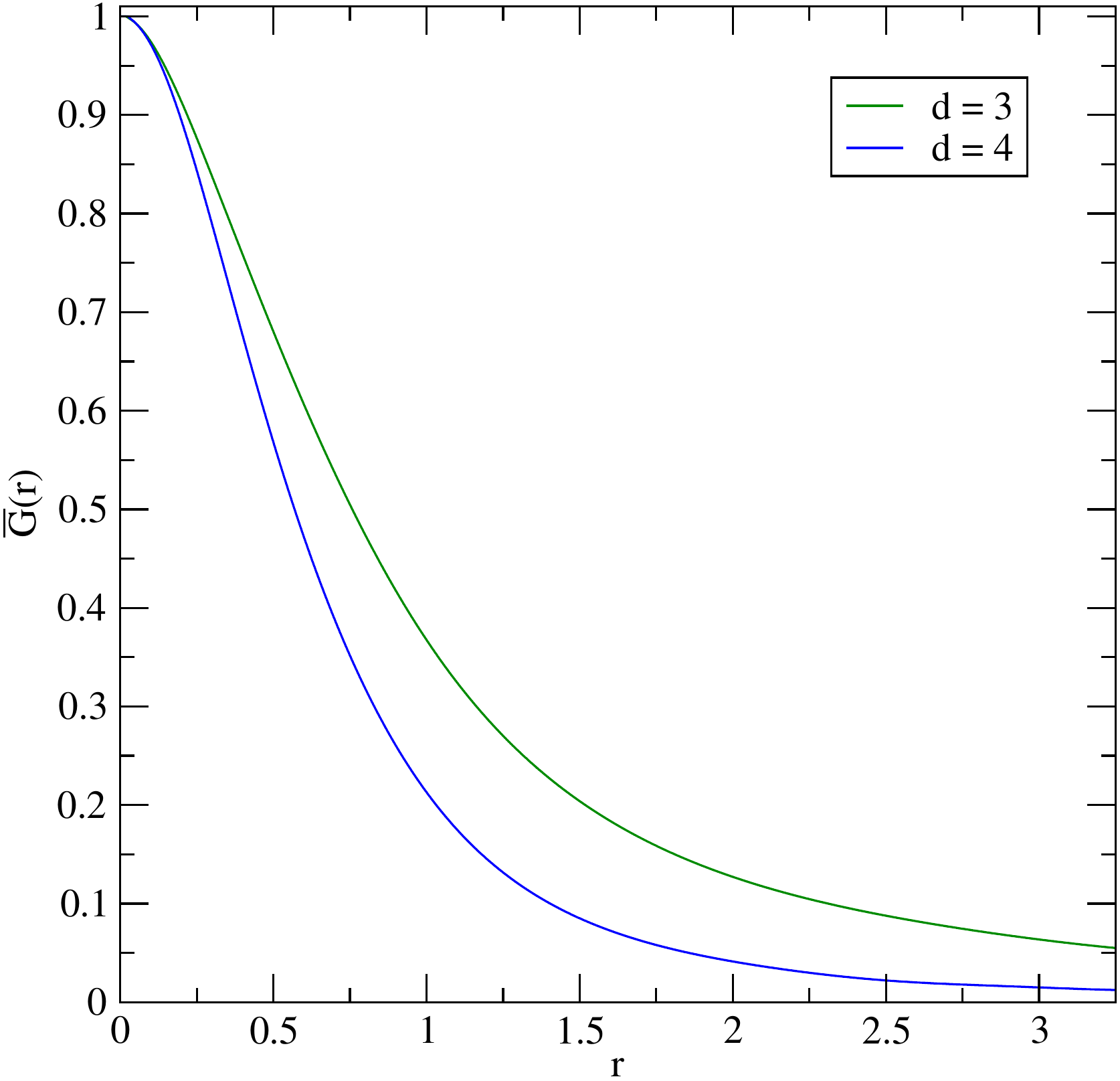}
\caption{Plots of $\tilde{G}(r) = G_V(r) - G_P(r)$ for $d = 1, 2, 3$ and $4$.}\label{Gtilde}
\end{figure}

We show in an accompanying paper \cite{ToScZa08} that for any $d$, $H_P \sim r^{d+1}$ for small $r$, and we observe this behavior in our results.  It is also true 
\cite{ToScZa08} that $H_V \sim r^{d-1}$, $E_V \sim 1-\chi(d)r^d$, and $E_P \sim 1-\varkappa(d)r^{d+2}$ as $r\rightarrow 0$,
where $\chi(d)$ and $\varkappa(d)$ are dimensionally-dependent constants.  
These properties imply that $G_P \sim r^2$ and $G_V \sim 1$
for small $r$ as with the $d = 1$ case.  Figure \ref{2dGpGv} shows these trends in greater detail.  
With regard to the large-$r$ behavior of $G_P$ and $G_V$, the linearity of both curves apparently holds in each dimension.  A surprising detail, however, is that 
$G_P$ and $G_V$ appear to converge with respect to increasing dimension.  
To understand this observation, we recall from \eqref{GpGvrel} that $G_P = G_V - \tilde{G}$; Figure \ref{Gtilde} provides 
plots of $\tilde{G}(r)$ for $d = 1, 2, 3$, and $4$.
It is clear from these curves that in each dimension $\tilde{G}$ for large $r$ is positive and scales more slowly that $r$ in each dimension.  We therefore expect that
the large-$r$ slope of $G_P$ is equal to the asymptotic slope of $G_V$ according to \eqref{GpGvrel}.  Since numerical results for $G_V$ are more easily and more
accurately obtained, we assume this asymptotic convergence and provide results for the asymptotic slope of $G_V$ below.
Table \ref{tableone} collects our calculations for the slope of $G_V$ in each dimension for large $r$.  The slopes are calculated by fitting 
the large-$r$ portion of each quantity
to a function of the form:
\begin{equation}
\digamma(x) = a_0 x + a_1 + \sum_{i=2}^n a_i \left(\frac{1}{x}\right)^{i-1}.
\end{equation}

It has been conjectured in \cite{ToScZa08} that as the dimension $d$ (finite) of the system increases, 
the asymptotic slope of $G_V$ and $G_P$ should approach the corresponding
value for a Poisson point process in dimension $d+1$.  The results in Table \ref{tableone} indicate that this claim closely holds for $d = 3$ and $4$, meaning that
the convergence of processes is relatively quick with respect to increasing dimension.  Based on the analysis in \cite{ToScZa08}, we therefore expect
this trend to continue for higher dimensions.

\begin{table*}[!htp]
\begin{tabular}{l c c}
\hline
\hline
$d$ &~~~~~& $G_V$\\
\hline
1 &~~~~~& $\pi^2/2$ (exact) \\
2 &~~~~~& $2.499\pm 0.015$ \\
3 &~~~~~& $1.680\pm 0.025$ \\
4 &~~~~~& $1.323\pm 0.049$ \\
\hline
\hline
\end{tabular}
\caption{Large-$r$ slopes of $G_V$ for each dimension.  The $d = 1$ slope is taken from the asymptotic expansion in \eqref{Gvlarges}.
Given errors are estimated based on the approximate error for $d = 1$.}\label{tableone}
\end{table*}

\subsubsection{Voronoi statistics of the Fermi sphere point process for $d = 2$}

\begin{figure}[!htbp]
\centering
\includegraphics[width=0.4\textwidth]{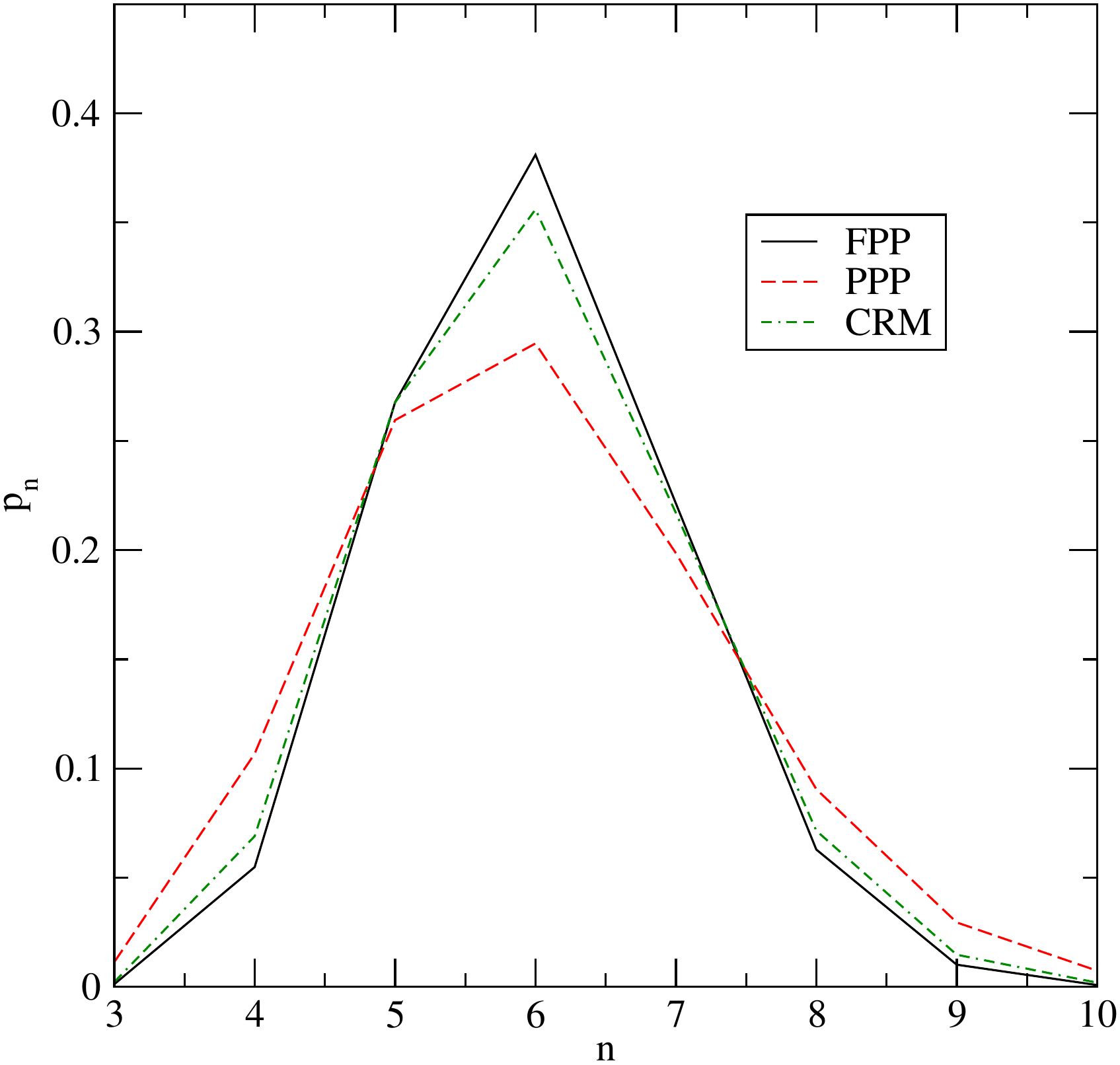}\hspace{0.5cm}
\includegraphics[width=0.4\textwidth]{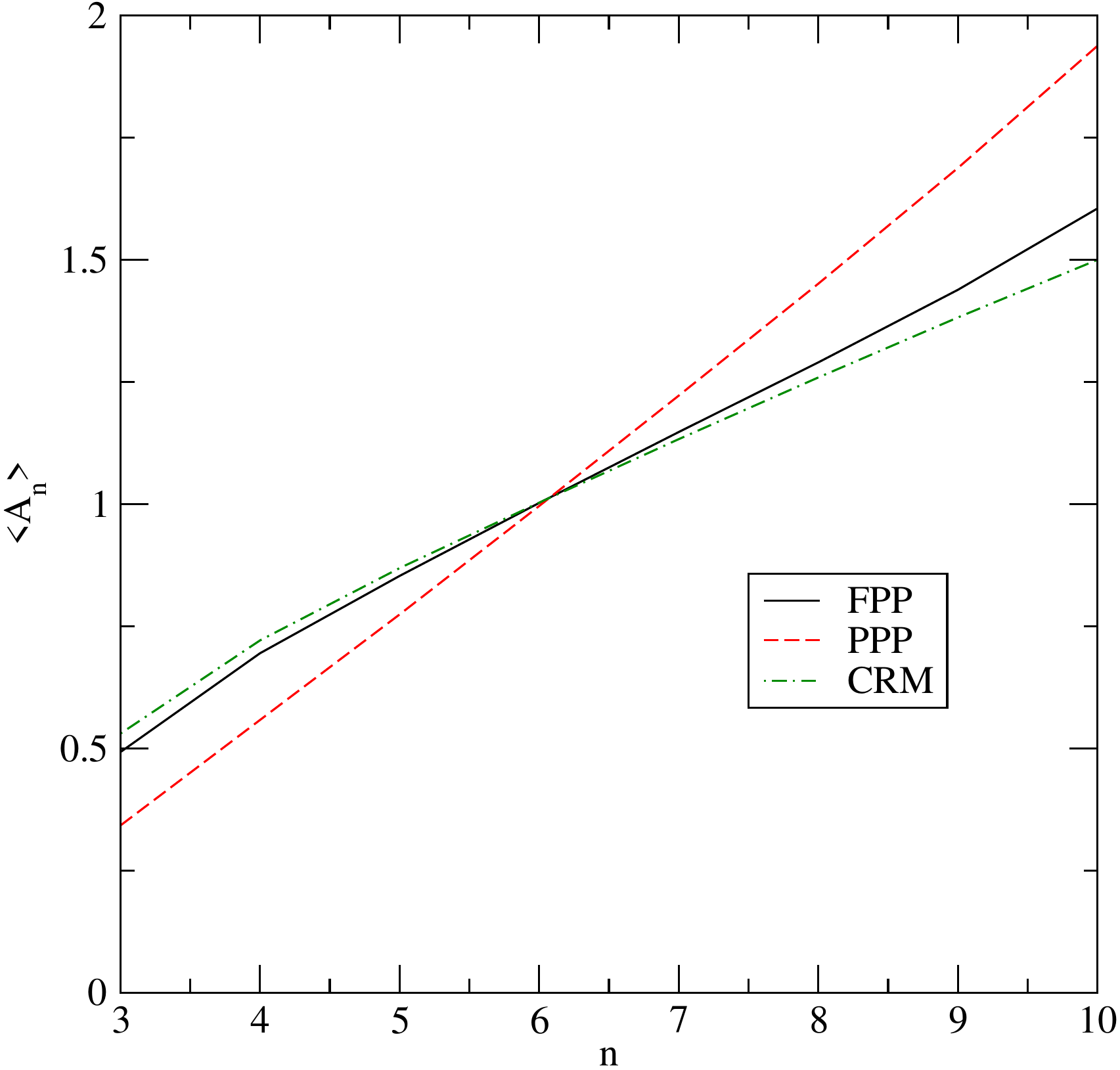}
\caption{\emph{Left:}  Distribution $p_n$ of the number of sides $n$ of Voronoi cells for the Fermi sphere point process (FPP), Poisson point process (PPP), 
and eigenvalues of a complex 
random matrix (CRM).\\
\emph{Right:}  Expectation value of the area of an $n$-sided Voronoi cell $\langle A_n\rangle$ for the FPP, PPP, and CRM.}\label{vorpnAnfig}
\end{figure}

To demonstrate the utility of the HKPV Algorithm in statistically characterizing a point process, we have also included statistics for the Voronoi tessellation 
of the $d = 2$ Fermi sphere point process in Table \ref{tablethree}.  Specifically, we provide results for the probability distribution of the number of cell sides $p_n$
and the average area of an $n$-sided cell $\langle A_n\rangle$.  Similar results have been reported in the literature for Voronoi tessellations of Poisson point processes
\cite{Hi05} and determinantal point processes generated from the eigenvalues of complex random matrices \cite{CaHo90}; we also provide the comparison in Table 
\ref{tablethree}.  Visual representations of the data are shown in Figure \ref{vorpnAnfig}.    

\begin{table*}[!htp]
\begin{ruledtabular}
\begin{tabular}{c|c c c c c c c c}
$n$ & 3 & 4 & 5 & 6 & 7 & 8 & 9 & 10\\
\hline
FPP; $p_n$ & 0.00124 & 0.05483 & 0.26770 & 0.38099 & 0.22136 & 0.06287 & 0.01013 & 0.00082 \\
PPP; $p_n$ & 0.0113 & 0.1068 & 0.2595 & 0.2946 & 0.1986 & 0.0905 & 0.0295 & 0.0074\\
CRM; $p_n$ & 0.0022 & 0.069 & 0.2676 & 0.356 & 0.217 & 0.0715 & 0.0147 & 0.0019\\
\hline
FPP; $\langle A_n\rangle$ & 0.49229 & 0.69469 & 0.85291 & 1.0024 & 1.1474 & 1.2900 & 1.4385 & 1.6051\\
PPP; $\langle A_n\rangle$ & 0.342 & 0.558 & 0.774 & 0.996 & 1.222 & 1.451 & 1.688 & 1.938 \\
CRM; $\langle A_n\rangle$ & 0.53 & 0.721 & 0.869 & 1.003 & 1.133 & 1.259 & 1.382 & 1.50\\
\end{tabular}
\end{ruledtabular}
\caption{Voronoi statistics for several point processes with $d = 2$. FPP = Fermi sphere point process; PPP = Poisson point process; CRM = complex random matrix.
Results for the PPP and CRM are from \cite{CaHo90}.  The systems have been normalized to unit number density ($\rho = 1$).}\label{tablethree}
\end{table*}

\begin{figure}[!htbp]
\centering
\includegraphics[width=0.4\textwidth]{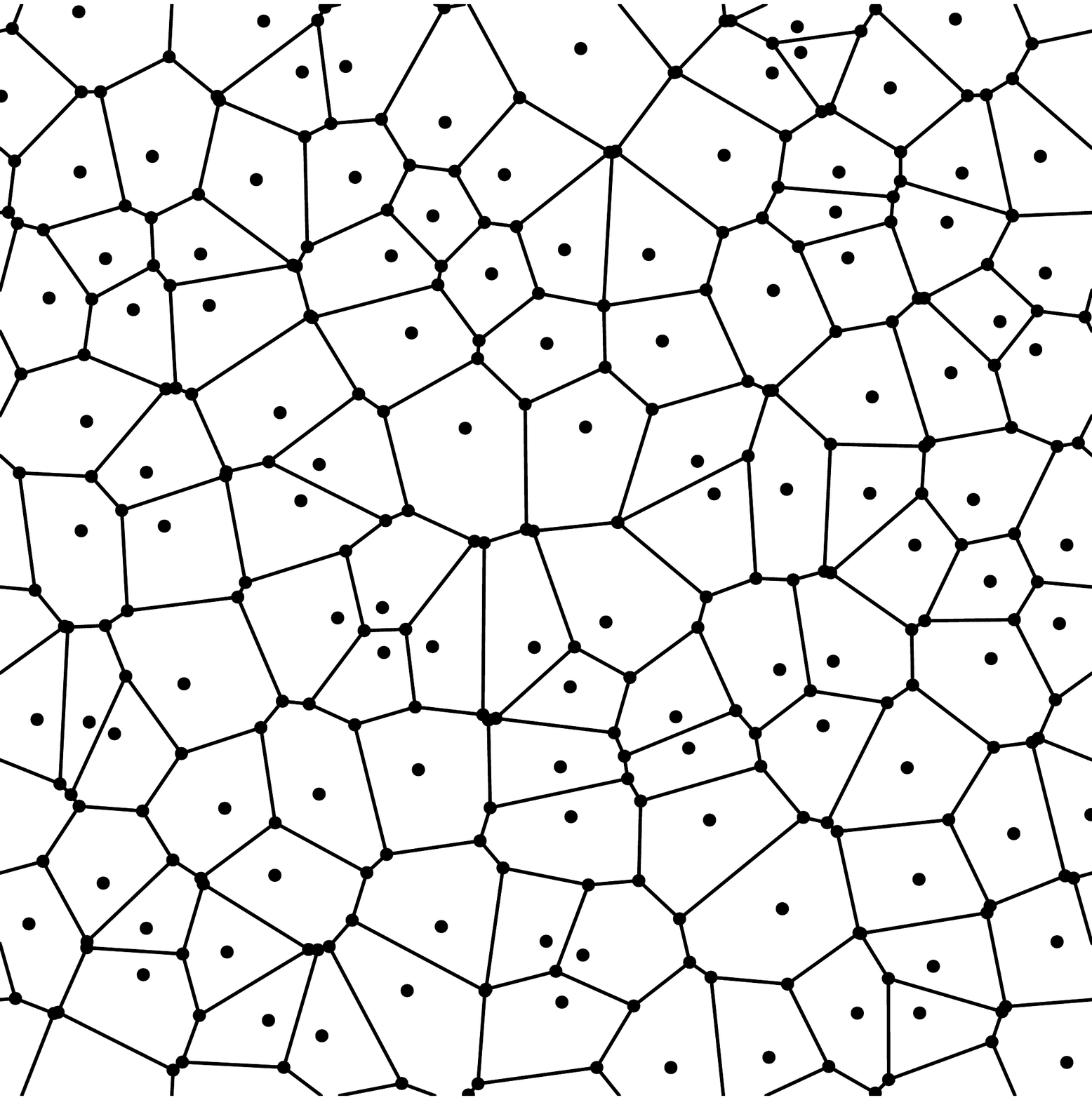}\hspace{0.05\textwidth}
\includegraphics[width=0.4\textwidth]{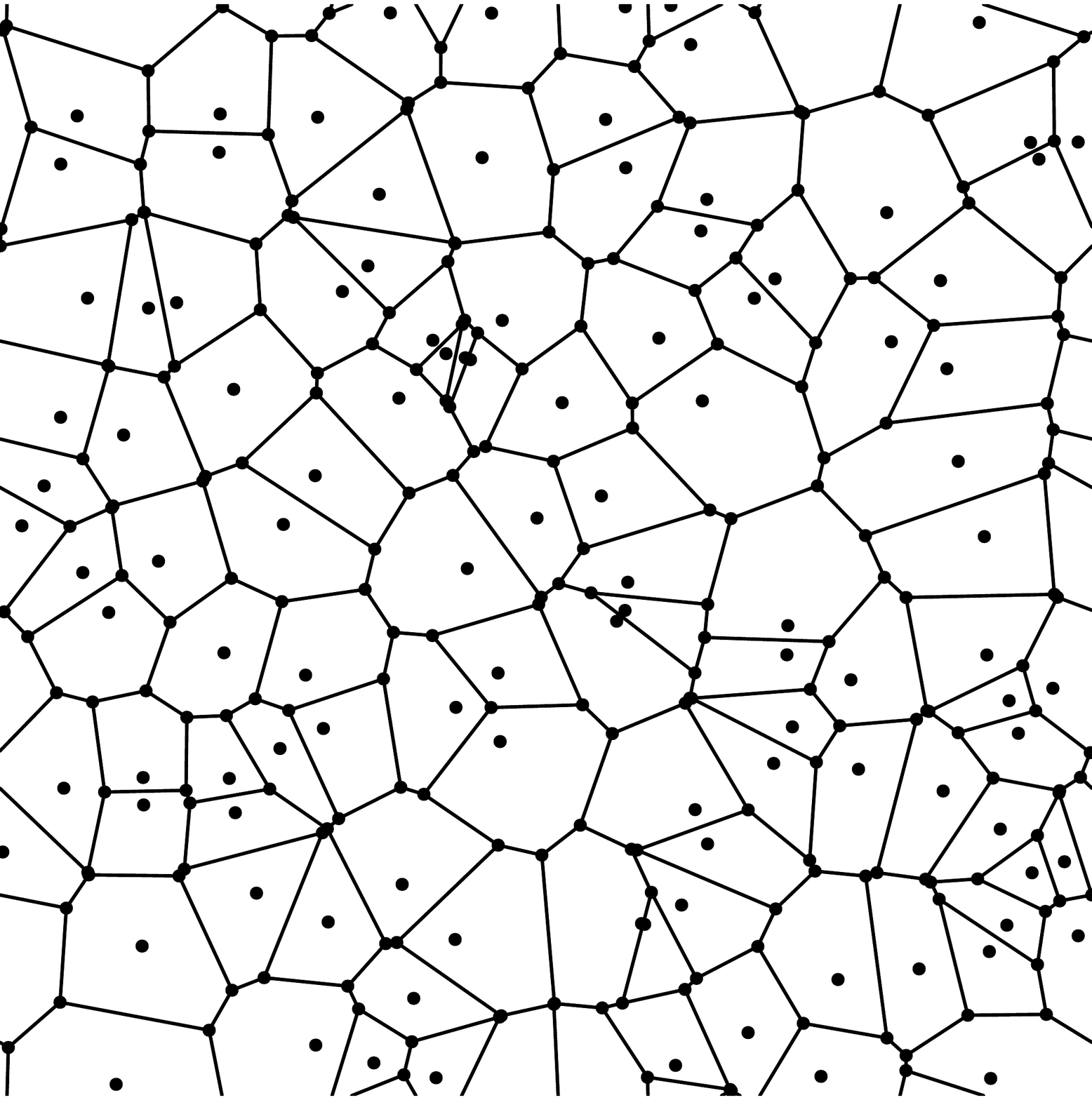}
\caption{\emph{Left:}  Voronoi tessellation of the $d = 2$ Fermi sphere point process at number density $\rho = 1$.\\
\emph{Right:}  Voronoi tessellation of a $d = 2$ Poisson point process at number density $\rho = 1$.\\
\emph{Both:}  Tessellations are performed with periodic boundary conditions using $N = 109$ points.}\label{vor}
\end{figure}

The topology of the plane enforces the constraints that $\langle n\rangle = 6$ and $\langle A\rangle = 1/\rho$ 
($=1$ at unit density) for any point process, where $n$ is the number of cell sides and $A$ is the area of a cell.  
We notice that the distribution $p_n$ is more sharply peaked for the Fermi sphere point process than in the Poisson point process, which is a consequence of
the effective repulsion among the particles.  With regard to the average areas of cells, is appears that Fermi-sphere cells with  smaller $n$ have larger areas than Poisson cells, again likely due to the repulsion of the points; however, Poisson cells with a greater number of sides tend to have larger areas than Fermi-sphere cells, a result
which can be attributed to the more even distribution of points in the Fermi-sphere process through space, which is related to the hyperuniformity of the point process.  
Figure \ref{vor} shows a typical Voronoi tessellation for the Fermi-sphere point process compared to the equivalent tessellation for a Poisson point
process. We immediately notice that the determinantal point process tends to avoid clustering of particles, resulting in a narrower distribution of cell
sizes within the tessellation; such clustering is not precluded in the Poisson tessellation, resulting in isolated regions of small (or large) cells.   

In order to rationalize these properties, we utilize the hyperuniformity (superhomogeneity) of the Fermi sphere point process.
Voronoi tessellations of hyperuniform point processes share several unique characteristics which distinguish them from general point processes.
For example, Gabrielli and Torquato \cite{GaTo04} have provided the following summation rule, which holds for all hyperuniform 
point processes in any dimension:
\begin{equation}\label{hypersumrule}
\lim_{V\rightarrow \infty}\left\langle\sum_{j=1}^{N(\mathcal{S})}w_iw_j\right\rangle = \sum_{j=-\infty}^{+\infty} C_{ij} = 0,
\end{equation}
where $V$ is the system volume, $N(\mathcal{S})$ is the number of points in a large subset $\mathcal{S}$ of $V$, $w_i = v_i-1/\rho$, $v_i$ is the size of 
Voronoi cell $i$, and $C_{ij} = \langle w_i w_j\rangle$ defines the correlation matrix between the sizes of different Voronoi cells. 
We note that this rule is essentially a discretization of the condition that $S(0) = 0$ for a hyperuniform point process, meaning that infinite-wavelength
number fluctuations vanish within the system; therefore, the result in \eqref{hypersumrule} is \emph{unique} to tessellations of hyperuniform
point processes.
Additionally, Gabrielli and Torquato have shown that arbitrarily large Voronoi cells or cavities are permitted in hyperuniform point processes despite the fact that these processes possess the slowest growth of the local-density fluctuation with $R$ (the size of the window)  \cite{GaTo04}.  We particularly emphasize the result that
the probability distribution of the void regions must decay faster in $R$ than the equivalent distribution for any non-hyperuniform process.  

We show elsewhere \cite{ToScZa08} that the structure factor in any dimension $d$
for the Fermi sphere point process has the following nonanalytic behavior at the origin:
\begin{equation}
S(k) \sim k \qquad (k\rightarrow 0),
\end{equation}
and the large-$R$ number variance is controlled by:
\begin{equation}\label{fermisigma}
\sigma^2(R) \sim R^{d-1}\ln(R).
\end{equation}
The unusual asymptotic
scaling $\sigma^2(R)/R^{d-1} = \ln(R)$ for the Fermi sphere point process has also been observed in three-dimensional maximally random jammed sphere packings
\cite{DoStTo05}, which can be viewed as prototypical glasses since they are both perfectly rigid mechanically and maximally disordered.

The peaking phenomenon observed in the Voronoi statistics of the Fermi sphere point process therefore reflects the fact that the probability of 
observing large Voronoi cells must be less than the corresponding probability for the Poisson point process, which is not hyperuniform.  
The more even distribution of the Voronoi cells through space in the Fermi sphere point process prevents the probability distribution
of the cell sizes from decaying more slowly than the corresponding distribution for the Poisson point process, where clustering of the points
increases the likelihood of observing both smaller and larger Voronoi cells.  

The comparison between the Voronoi statistics of the Fermi sphere point process and the Ginibre ensemble in Figure \ref{vorpnAnfig} highlights the similarities
between the two determinantal point processes.  Namely, the distributions $p_n$ for each system are sharply peaked around $n = 6$ and narrower than the 
corresponding result for the Poisson point process. However, notable differences between the statistics are also apparent.  The distribution $p_n$ for the 
Fermi sphere point process is more sharply peaked than the corresponding result for the Ginibre ensemble.  The larger probability in the Ginibre
ensemble of observing cells with a fewer or larger number of sides $n$ is directly related to the correlations among the particles in the system.

\subsection{Comparison of results across dimensions}

In order to compare statistical quantities across dimensions, it is generally preferable to enforce a \emph{fixed mean nearest-neighbor separation} $\lambda$
since this quantity determines the length scale of the system.  This constraint is easily obtained via a rescaling of the density according to the relation:
\begin{equation}\label{nndensity}
\lambda(\rho) = \lambda(1)\left(\frac{1}{\rho}\right)^{1/d},
\end{equation}
where $\lambda(1)$ denotes the mean nearest-neighbor separation at unit density.  Equation \eqref{nndensity} easily follows from the scaling of the density $\rho$ 
with the size of the system. Of particular interest are the values of $\lambda(1)$ for each dimension and $\rho(1)$, the number density at which the system has unit mean nearest-neighbor separation.
These quantities may be read from Table \ref{tabletwo}. 

It is not difficult to show, using \eqref{nndensity}, that $\rho(1) = \lambda(1)^d$.  We note \ASd{from Figure \ref{nnrho}} that for sufficiently large $\rho$, the mean
nearest-neighbor separation increases with the dimension of the system; however, the opposite trend is observed for small $\rho$.  For intermediate values of the density, 
the trend becomes less discernable.  At unit density, we observe that $\lambda(1)$ decreases between $d = 1$ and $d = 2$
but then increases again for $d \geq 2$; indeed, we measure this trend directly in Table \ref{tabletwo}.  Estimates for $\lambda(1)$, which are 
developed elsewhere \cite{ToScZa08}, suggest that $\lambda(1)$ continues to increase with respect to increasing dimension; if this result is true, then we
therefore expect that as $d\rightarrow \infty$, $\rho(1) \rightarrow \infty$.  From the definitions of $g_2$ and $k_F$  in \eqref{fermiong2}, one can 
show that $g_2(r) = g_2^{(1)}[\lambda(1)r]$, where $g_2^{(1)}$ is the form of the pair correlation function at unit density.  Therefore, as $\lambda(1)$ increases,
the curve representing $g_2$ shifts to the left, implying that for large dimensions $g_2$ is approximately given by unity for all $r$, and the system is uncorrelated.  
This behavior is a direct consequence of enforcing a fixed mean nearest-neighbor separation on the system as opposed to a fixed density.      

\begin{table*}[!htp]
\begin{ruledtabular}
\begin{tabular}{c c c}
$d$ & $\lambda(1)$ & $\rho(1)$\\
\hline
1 & 0.725728 & 0.725728 \\
2 & 0.649823 & 0.422270 \\
3 & 0.654511 & 0.280382 \\
4 & 0.679561 & 0.213262 \\
\end{tabular}
\end{ruledtabular}
\caption{Values of $\lambda(1)$ and $\rho(1)$ for each dimension.}\label{tabletwo}
\end{table*}

\begin{figure}[!htbp]
\centering
\includegraphics[width=0.4\textwidth]{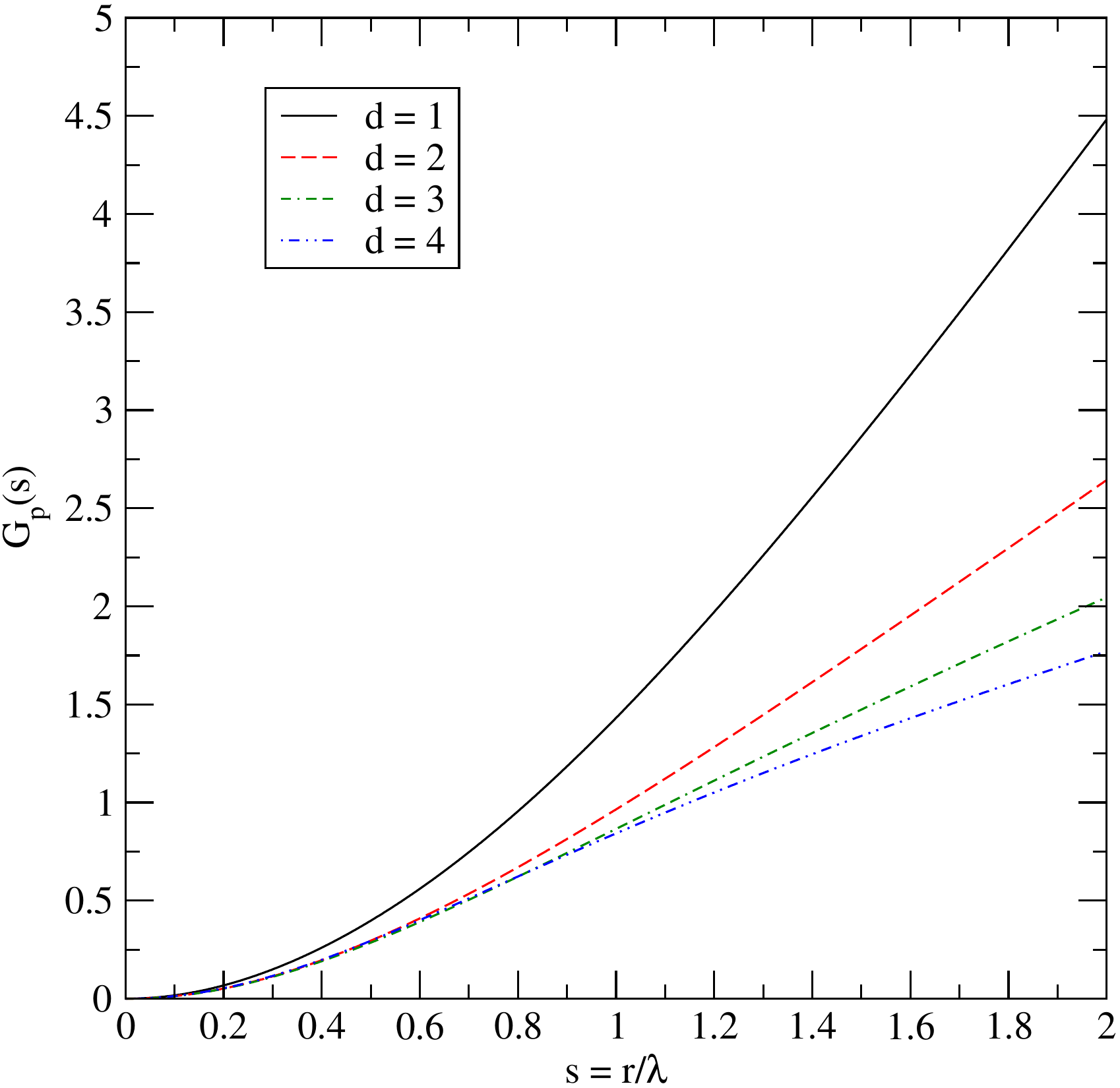}\hspace{0.5cm}
\includegraphics[width=0.4\textwidth]{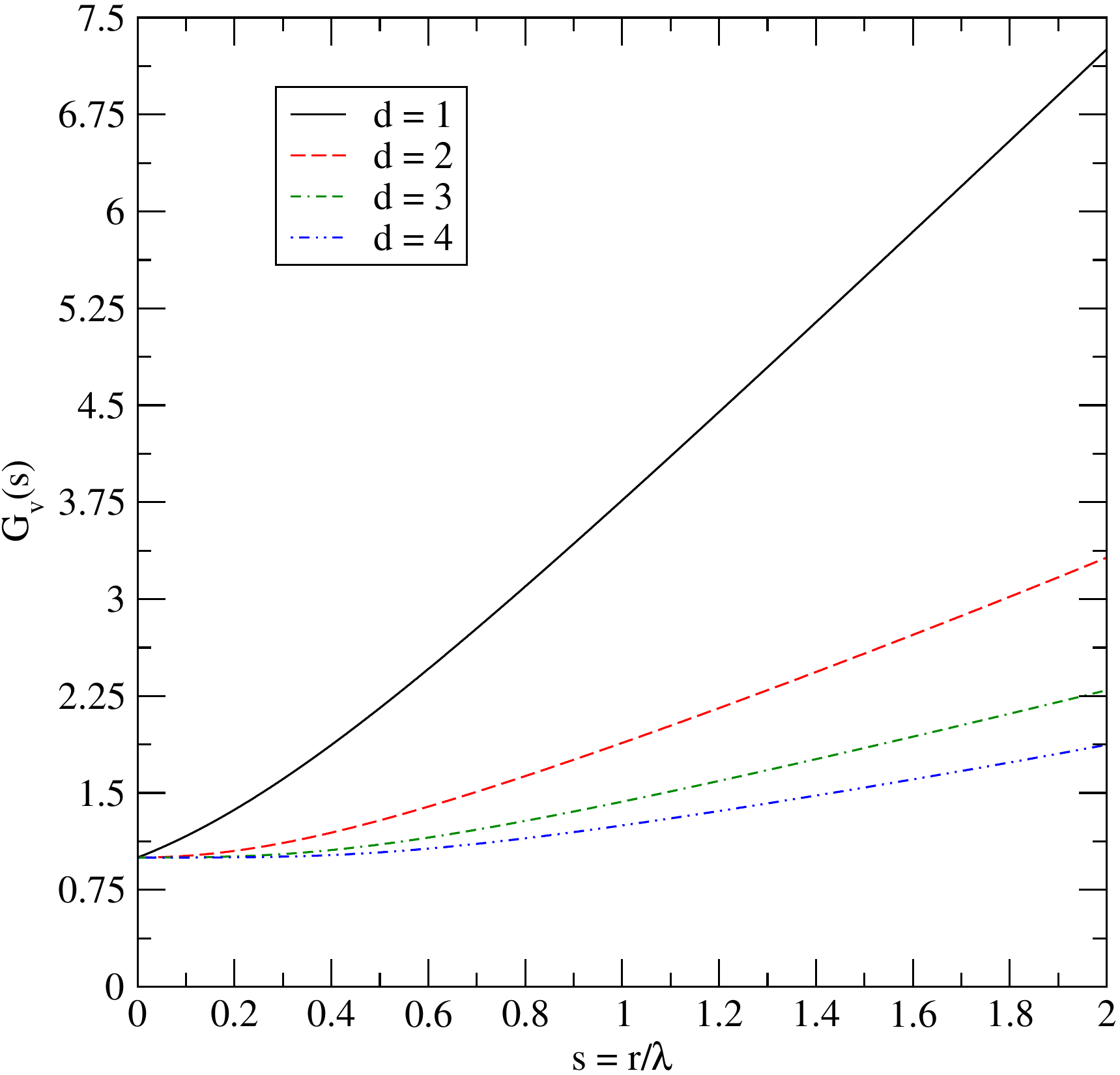}
\caption{Comparison of $G_P$ (\emph{left}) and $G_V$ (\emph{right}) across dimensions at unit mean nearest-neighbor separation $\lambda$.  Results are from 
numerical calculations using \eqref{eq:PD0M}. }\label{alldGpGv}
\end{figure}

After appropriate rescaling, we compare the results for $G_P$ and $G_V$ in Figure \ref{alldGpGv}.  The results strongly suggest that $G_V(r) \rightarrow 1$ as 
$d\rightarrow \infty$, which is in agreement with the conclusions drawn from the analysis above.  We also notice that both $G_P$ and $G_V$ decrease in slope
as the dimension of the system increases; thus, if $G_P$ and $G_V$ possess the same $r\to\infty$ asymptotic slope, then it must be true that $G_P$ saturates
at unity for large $r$ in the limit $d\rightarrow \infty$. This behavior is surprising in the context of our description of $g_2$ above.  The fact that 
$g_2 \rightarrow 1$ for large $d$ indicates a decorrelation of the system for higher-dimensions, leading us to expect Poisson-like behavior in the system
as conjectured in \cite{ToScZa08}.  
The behavior of $G_V$ corroborates this notion as does the convergence of $G_P$ and $G_V$ for large $d$.  However, our understanding of $H_P$ and $E_P$ from 
the discussion above along with the bounds from \cite{ToScZa08}, which sharpen with increasing dimension at fixed $\lambda$, 
suggest instead that $H_P\rightarrow H_P^* = \delta(r-1)$ and $E_P\rightarrow E_P^* = \Theta(1-r)$ for large $d$,
where $\delta(x)$ is the Dirac delta function, $\Theta(x)$ is the Heaviside step function, and 
$H_P^*$ and $E_P^*$ are effective generalized functions.  
As shown in \cite{ToScZa08}, the only functional form for $G_P$ that agrees with 
these conclusions and the observed behavior in Figure \ref{alldGpGv} is $G_P \rightarrow G_P^* = \Theta(r-1)$ as $d \rightarrow \infty$ for fixed 
mean nearest-neighbor separation.   

\begin{figure}[!htp]
\centering
\includegraphics[width=0.4\textwidth]{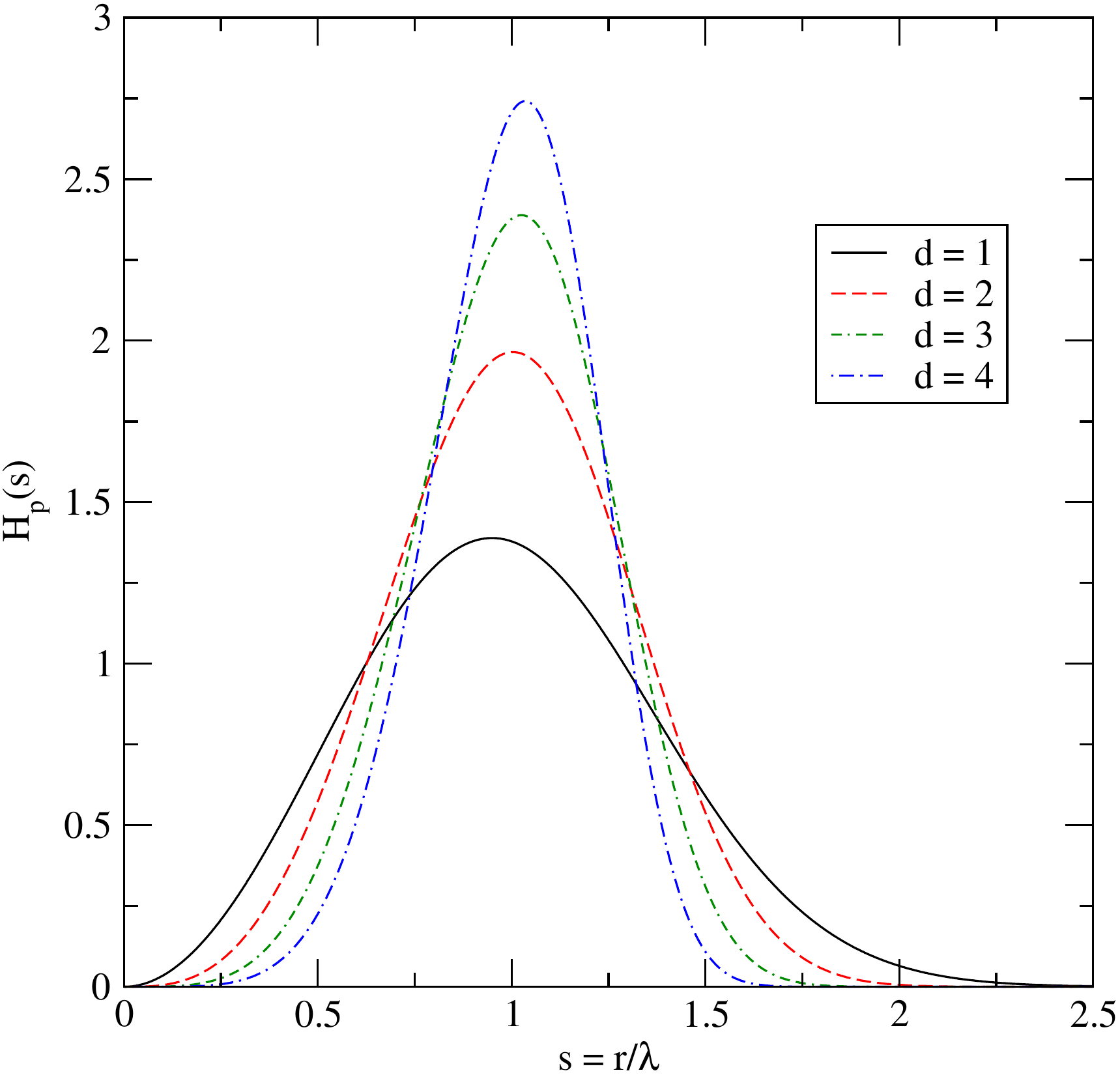}\hspace{0.5cm}
\includegraphics[width=0.4\textwidth]{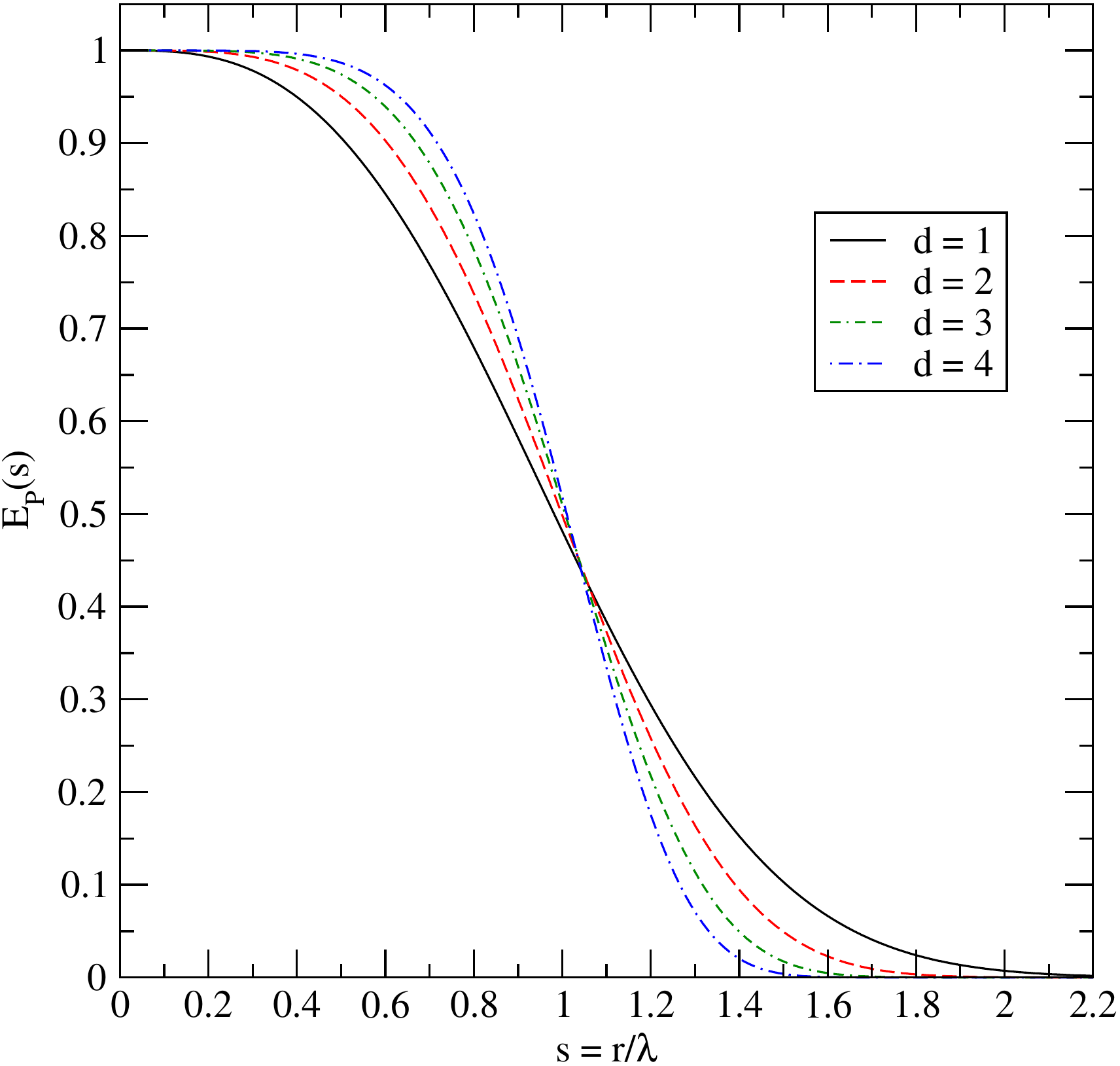}
\caption{\emph{Left:}  $H_P(s)$ for the Fermi sphere point process at unit mean nearest-neighbor separation $\lambda$ for $d = 1, 2, 3, 4$. \emph{Right:}  $E_P(s)$ for the Fermi sphere point process at unit mean nearest-neighbor separation $\lambda$ for $d = 1, 2, 3, 4$.}\label{alldHpEp}
\end{figure}

We rationalize these observations by noting that the effective hard core of the fermionic system as described in \cite{ToScZa08} 
has been encoded in the functional form of $G_P$
due to the constraint of fixed mean nearest-neighbor separation.  It is this constraint which produces the limiting forms of $H_P$ and $E_P$ for high dimensions,
meaning that the environment around any given particle greatly resembles a saturated system of hard spheres.  However, the scaling of $g_2$ with $\lambda(1)$ 
mentioned above means also that the particles only see large-$r$ correlations from the corresponding form of $g_2$ at unit density, resulting in
Poisson-like behavior for this function \cite{ToScZa08}, which is then translated into the value of unity for $G_V$ in high dimensions.  In other words, the 
particle quantities contain the information about the effective hard core under the constraint of 
fixed mean nearest-neighbor separation, but the void quantities are Poisson-like to account both for the scaling of $g_2$ and the small- and large-$r$
contraints shown numerically in Section IV.B that must be enforced regardless of how the infinite-dimensional limit is taken.  

\begin{figure}[!htbp]
\centering
\includegraphics[width=0.41\textwidth]{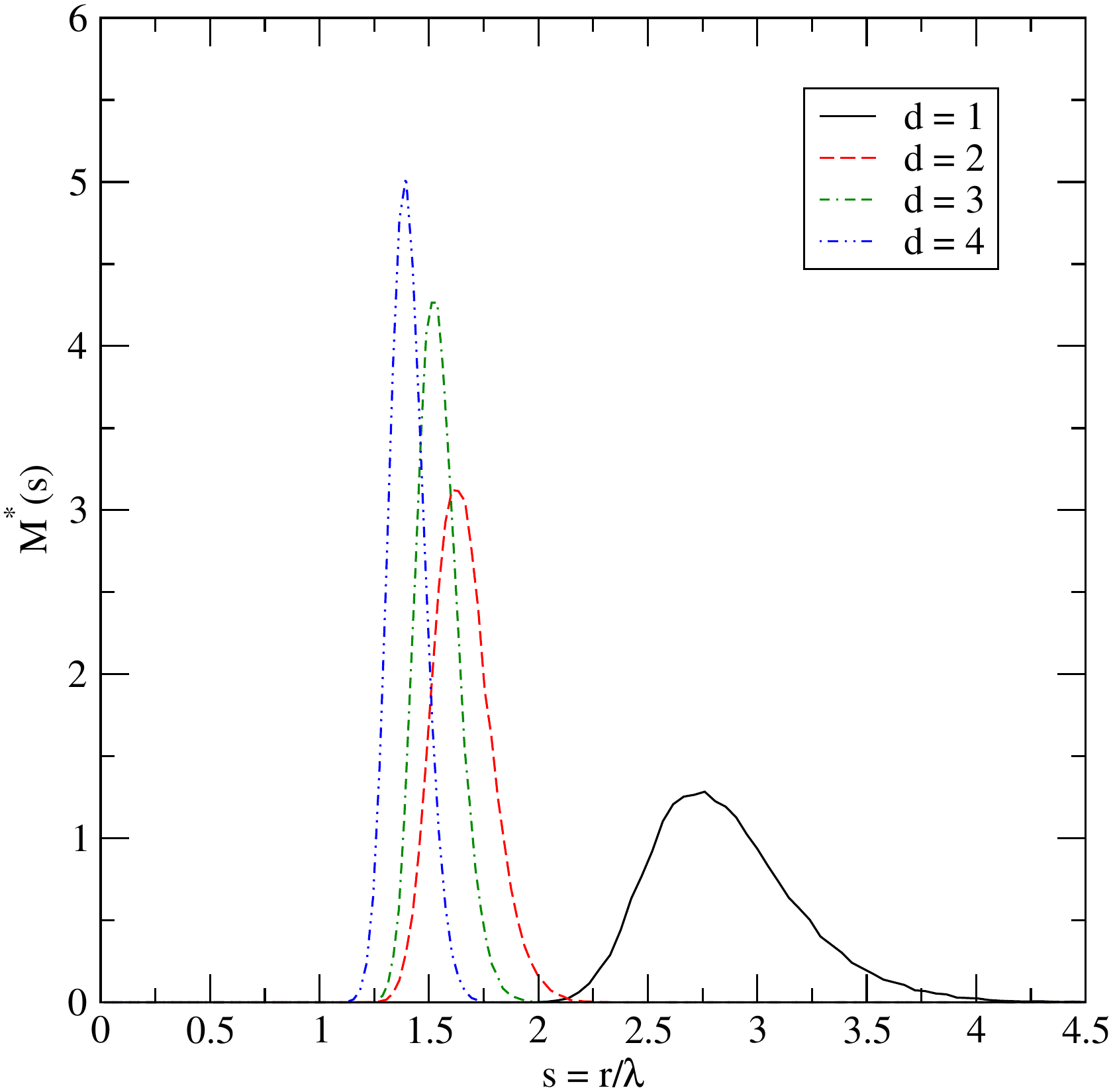}\hspace{0.5cm}
\includegraphics[width=0.4\textwidth]{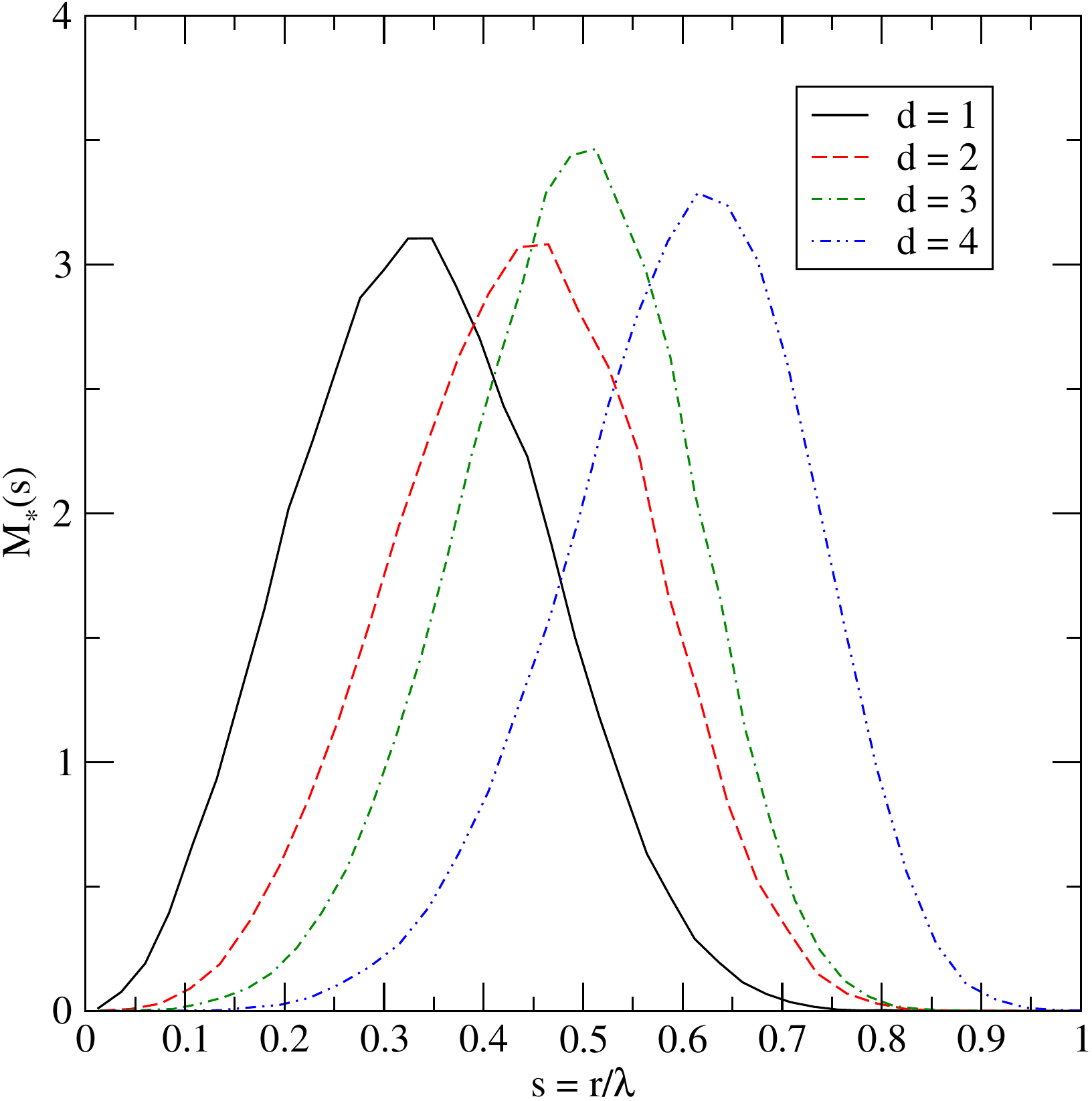}
\caption{Distributions of the maximum ($M^*(s)$; \emph{left}) and minimum ($M_*(s)$; \emph{right}) 
nearest-neighbor distances across dimensions at unit mean nearest-neighbor separation $\lambda$.
Results are simulated using the HKPV Algorithm.}\label{alldmaxmin}
\end{figure}

\begin{figure}[!htbp]
\centering
\includegraphics[width=0.45\textwidth]{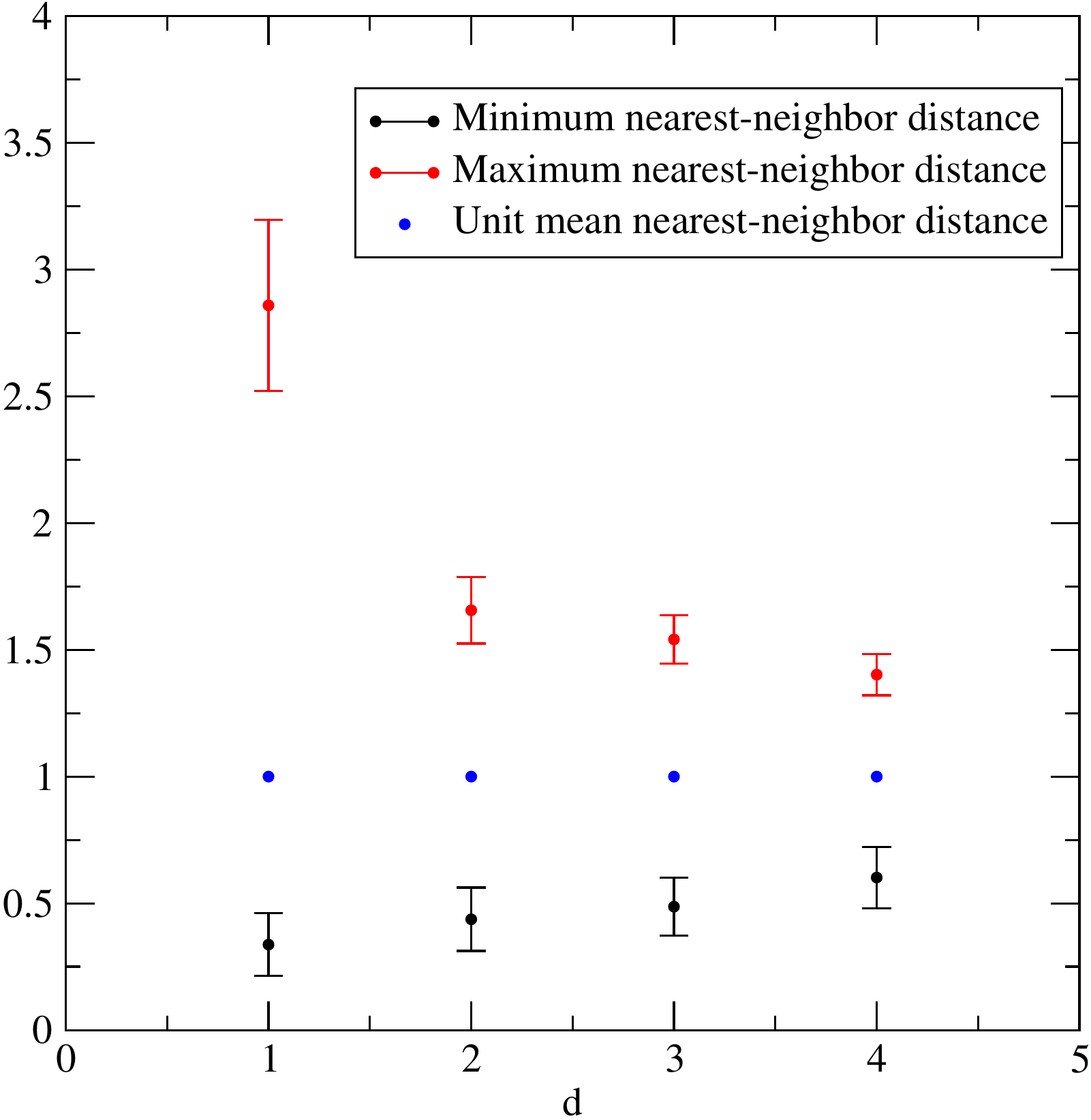}
\caption{Average maxima and minima nearest-neighbor distances with standard deviations across dimensions at unit mean nearest-neighbor separation;
results are obtained using the HKPV Algorithm.  
Also included for reference is the unit mean nearest-neighbor separation, which is fixed for each dimension.}\label{alldmaxminmean}
\end{figure}

Figure \ref{alldmaxmin} shows the distributions of the extremum nearest-neighbor distances at fixed mean nearest-neighbor separation based on 
calculations from configurations generated with the HKPV Algorithm.  We have been unable to write these quantities in determinantal form
amiable to numerical calculation, and therefore the HKPV Algorithm is an attractive means through which to study these quantities.  We note that the maximum and minimum
nearest-neighbor spacings appear to converge to a value of unity as the dimension of the system increases; this behavior is expected in the context of the 
discussion for $H_P$ above.  The convergence of these quantities is more easily seen in Figure \ref{alldmaxminmean}; we have also included the values of 
$\lambda(1)$ for reference, but there is strong evidence to suggest that the limiting value of the extremum quantities for large $d$ is unity.

\section{Determinantal processes in curved spaces}

\begin{figure}[!htbp]
\centering
\includegraphics[width=0.45\textwidth]{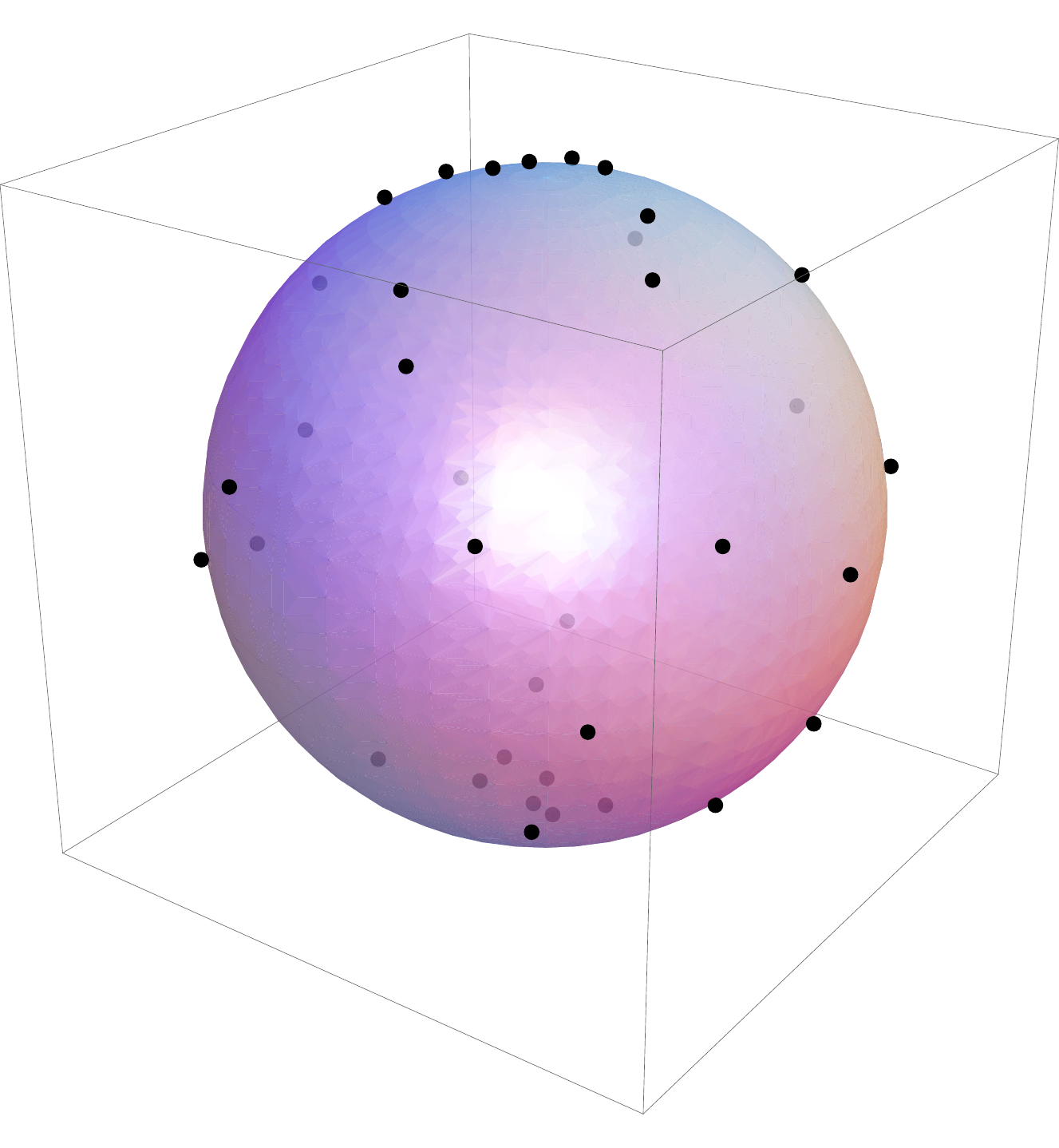}
\caption{Configuration of 37 points on the unit sphere using the HKPB algorithm with the spherical harmonics as basis functions.}\label{sphharm}
\end{figure}

In this last section, we present an example of how the HKPB algorithm is not limited to point processes in Euclidean spaces described above. With an appropriate choice of the basis functions $\phi_n$ it  can in principle be adapted to simulate point processes on other domains and topologies. Of particular interest in this regard is the 
generation of point processes on a curved space, like the two-sphere $S^2$ in Figure \ref{sphharm}.  
Here, we consider the spherical harmonics as basis functions for a spherical geometry; $\phi_n=Y_{l,m}(\theta,\phi)$ are a basis for the square-integrable functions on the two-sphere $S^2$. Since $m=-l,...,l$ so for any $l$ there are $2l+1$ different values of  $m$, we decided to choose the lowest $(N-1)/2$ values of $l$ and all the corresponding $m$'s. Once these functions have been chosen, the algorithm provides a relatively simple means to generate the point process. We have not embarked in an extensive analysis of the statistical properties of this process as we leave that for future work. We note, however, from previous observations that a short-distance effective interaction among the points is logarithmic and repulsive, and we expect a fluid-like configuration on the surface of the sphere. Also, for $N\to\infty$ at fixed sphere radius it is not difficult to conjecture that the nearest neighbor functions will tend to those we already discussed for the Fermi-sphere process on torus. On the other hand, for finite $N$ this problem could be relevant to problem of packing of spheres in non-Euclidean geometries. This is a promising direction for future research.

\section{Concluding Remarks}

Our focus in this paper has been on characterizing the statistical properties of high-dimensional determinantal point processes through both 
numerical calculations and algorithmic generations of point configurations. We first compared the results for $n$-particles distribution functions and nearest-neighbor function obtained by the two methods to cross-check consistency and accuracy. We then proceeded using both methods to elucidate the small- and large-$r$ behaviors of the nearest-neighbor distribution functions and the extrema statistics in dimensions one to four. Our results strongly suggest that both $G_P$ and $G_V$ \emph{are linear} for sufficiently large $r$, and we obtain numerical estimates of the common slope in this limit.  This behavior is to be contrasted with the equivalent forms of $G_P$ and $G_V$ for 
equilibrium systems of hard spheres and for Poisson point processes. It is known for the former system that both functions saturate for sufficiently large $r$ while 
$G_P(r) = G_V(r) = 1$ for all $r$ in the latter process \cite{torquato2002rhm}.  
The linearity of $G_P$ and $G_V$ in the determinantal case is thus unique in the context of general point processes.  We have also shown, in accordance with \cite{ToScZa08}, that in the limit as $d \rightarrow \infty$, both $G_P$ and $G_V$ must saturate at unity in accordance
with the behavior of the aforementioned bounds; again, this claim is supported by the numerical evidence we have presented here.  Also, as the dimension $d$ grows, we observed that the functions $H_P$ and $H_V$ become concentrated around their maximum as do the distributions of extrema of nearest-neighbor distances $M_*$ and $M^*$. 

By using the HKPV algorithm to generate configurations of points we have shown that the determinantal nature of the $n$-particle probability density has a significant effect on the Voronoi cell statistics of the Fermi-sphere point process in two dimensions. Namely, the probability distribution of cell-sides is more peaked around $n=6$ (hexagons) than the corresponding distribution for either Poisson point process or the Ginibre ensemble \cite{CaHo90} (the distribution of complex eigenvalues of random complex matrices). The effective separation of the points, resulting in a sharper peak in the distribution of the number of sides of the Voronoi cells, is closely related to the hyperuniformity of the system. 

Finally, to show how the algorithm can be used for generating determinantal processes on curved spaces, we have presented an example of determinantal point process on the two-sphere.

\bibliography{fermionnumpaper1028}

\end{document}